# Dimethylamino terminated ferroelectric nematogens revealing high permittivity

Martin Cigl, Natalia Podoliak, Tomáš Landovský, Dalibor Repček, Petr Kužel,

and Vladimíra Novotná*

*Institute of Physics of the Czech Academy of Sciences, Na Slovance 2, Prague, Czech Republic*

**Abstract**

Since the recent discoveries, ferroelectric nematics became of upmost interest due to their outstanding ferroelectric properties. In this work, we prepared a series of polar molecules revealing a ferroelectric nematic phase ($N_F$) with a very high dielectric constant (>104). A new motif, which differs from previously reported molecular structures, was optimized to support the $N_F$ phase. For all homologues the $N_F$ phase was observed directly on the cooling from the isotropic phase and ferroelectric behaviour was investigated by dielectric spectroscopy, second harmonic generation, polarization current measurements and by analysis of textures in the polarized light. The presented materials combine ferroelectricity with giant permittivity in a fluid media at room temperatures, so they appear to be extremely attractive. Polarity of molecules with the strong susceptibility to the electric field represent high potential for various applications in energy-efficient memory devices or capacitors.

## 1. Introduction

In thermotropic liquid crystals (LCs) molecules can self-assemble and create intermediate phases (mesophases) in a certain temperature range between liquid and crystalline phases [1], combining the fluidity of liquids with the anisotropy characteristic for crystals. Anisotropic properties of LC medium manifest itself as a result of the anisotropic shape of (partially) ordered constituent molecules. A large variety of phases and structures can be observed in LCs, which are susceptible to external field and boundary conditions. Many LC phases reveal a large electro-optical response, which became a background of mass-production technological applications (monitors, sensors, etc.). First, the ferroelectricity in LCs was associated with chirality of the constituent molecules and only a tilted smectic phase formed by chiral rod-like molecules [1] was considered to feature ferroelectricity (FE) and/or antiferroelectricity (AF). With the discovery of bent-core materials [2], it was found that non-chiral mesogens may also form FE and AF phases as the close packing and hindered rotation can lead to the structural chirality. Nevertheless, due to a higher viscosity, smectic phases never reached such broad application range as nematics.

Recent discoveries stimulated renewed intensive progress in the field of nematic liquid crystals. For a conventional nematic phase, the director orientations *n* and *–n* are indistinguishable due to the thermal fluctuations, so they form only non-polar phases. However, as far back as in 1918, Max Born [3] predicted a possibility of a ferroelectric fluid, in which all the dipoles point in the same direction. In such a nematic ferroelectric state ($N_F$), the dipole moments $\mu$ should be strong enough such that the dipole-dipole interactions overwhelm the thermal fluctuations. In 2017, a real breakthrough was announced in the development of LCs, as the first two ferroelectric nematics (denoted RM734 and DIO) were reported simultaneously by two research teams [4-6]. Both materials reveal extremely high longitudinal dipole moments (about 10 D), anomalously huge dielectric anisotropy $\Delta\varepsilon$, and a spontaneous polarisation of about 4 $\mu C/cm^2$, which is an order of magnitude higher than the previously reported values in other ferroelectric LC phases. Recently, these materials have been intensively studied [7-17]. Mandle at al. [9] synthesised a homologue series relevant to the molecular structure of RM734 and analysed the mesomorphic properties and tendencies leading to the $N_F$ phase. The compounds have been intensively studied by Ljubljana researchers [10-12] and by the Boulder group [13,14]. The existence of ferroelectric domains with a different macroscopic orientation of the dipoles in the absence of electric field was reported [10-14]. Details of polar nature of self-assembly, evolution of topological objects and analysis of their character [12,17,18] are under intensive research progress. Currently, the research is focused on the preparation and characterisation of new compounds. Machine learning procedures were applied to predict ideal conditions for the $N_F$ phase presence, including a dipole moment value, aspect ratio, length of the molecule as well as the dipolar angle [19]. In spite of the fact that these conditions are rather restrictive, development in the designing of prosperous molecular structures was promoted.

At the moment, microscopic organisation of the polar molecules and the mechanism of the phase transition to the ferroelectric nematic phase undergo intensive research and stimulating debates. A theoretical description of the ferroelectric nematic phase has been proposed [20,21], and chiral analogues of highly polar molecules were developed recently [22]. Additionally, a possibility of oligomer synthesis was shown [23] and new phases and effects introduced. In any case, the ferroelectric properties combined with the giant permittivity in a fluid media represent an attractive rapidly developing subject. Since the discovery of $N_F$ phase, the ongoing research is mostly concentrated on the design of new molecular structures. Up to now, the library of $N_F$ materials is strictly limited to a couple of general structures possessing a suitable aspect ratio and a large enough dipole moment, which develops due to the effective electron donating and withdrawing groups within the molecules.

In this contribution, we demonstrate newly designed structural motif (see Fig.1). In contrast to the previously reported molecular designs [3-5,8-19], which utilise an oxygen-based electron donating group, we synthesised a series possessing a more efficient nitrogen electron donating group in the terminal part of the aromatic system. Such a design yields higher dipole moment along the long molecular axis compared to other published materials. To modify the lateral interactions, which are strong in highly polar systems, we introduced a lateral alkyl chain with varied number of carbon atoms from 1 to 6. Based on these considerations, we synthesised a series of compounds (Fig. 1) which exhibit the $N_F$ phase directly below the isotropic liquid on cooling. By tuning the lateral substitution, we shifted the temperature interval of $N_F$ down to the room temperature (RT), at which it may eventually relax to a stable glassy state preserving the ferroelectric behaviour.

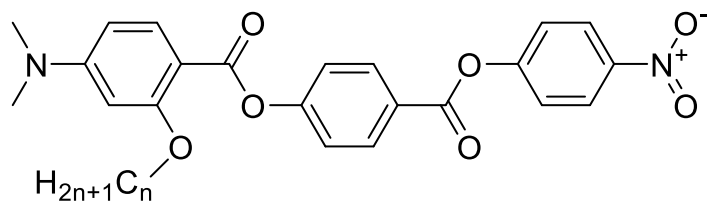

Fig. 1. Chemical formula of compounds **NFn** with n = 1 - 6.

## 2. Materials and methods

Chemical formula of the studied compounds is presented in Fig. 1. Synthesis of materials started from commercial 4-aminosalicylic acid (**1**, see Scheme 1). Its amino group was protected by acetylation and the carboxylic group was protected by alkylative esterification by methyl iodide, so as neither of the two groups interfere with the alkylation of phenolic hydroxyl. Protected derivative **2** was then alkylated by 1-bromoalkanes to get a series of alkyl homologues **3-n**. In the next steps, the acetyl group was cleaved by acidic hydrolysis under mild conditions and the liberated amino group was alkylated by dimethyl sulphate yielding the key intermediate, acid **4-n**. The lowest alkyl homologue (**4-1**) was synthesised directly from acid **1** by alkylation with the excess of dimethyl sulphate. The second part of the molecular core was synthesised from 4-hydroxybenzoic acid (**5**), which was protected by the reaction with 3,4-dihydro-2H-pyrane and reacted with 4-nitrophenol in a DCC-mediated esterification. The protected hydroxyl group was then liberated by the treatment with *p*-toluenesulfonic acid. The final step of the synthesis was esterification of acids **4-n** with phenol **6** mediated by EDC.

Differential scanning calorimetry (DSC) measurements were performed to acquire thermal properties. For electro-optical studies, a polarising optical microscope was used, equipped with a heating/cooling stage. Details about the compound characterisation and experimental apparatus are in Supplemental file.

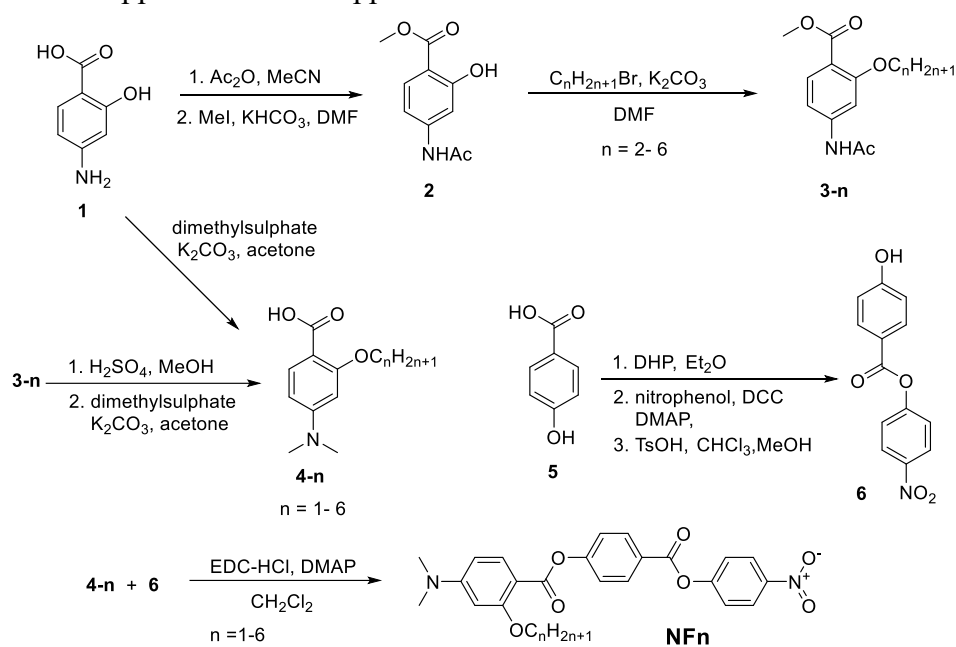

Scheme 1. Synthesis of the studied polar nematogens denoted **NFn** with *n* varying from 1 to 6.

## 3. Results

We studied all newly synthesised homologues by DSC and observed textures and their changes in polarising microscope to assess the phase behaviour. We performed DSC measurements in a broad temperature range. We established the melting point (m.p.) from the first heating run, during which we observed a direct transformation from the crystalline to the isotropic (Iso.) phase. After the first heating of the fresh sample, we followed with a cooling run from the Iso phase down to -25°C. On the cooling run, the compounds transformed to the liquid crystalline state at a significantly lower temperature, $T_{iso}$. Under the polarising microscope, we observed characteristic textures in the LC state, which were previously ascribed to the ferroelectric nematic phase, $N_F$ [12-18]. In the following description, the properties of $N_F$ phase are systematically uncovered.

The analysed DSC data are summarised in Table 1. Compounds **NF1**, **NF2**, and **NF3** did not crystallise during the cooling run, however, they crystallised during the subsequent heating. These homologues revealed the ferroelectric nematic phase only during the cooling of the sample; temperature stabilisation or heating of the sample caused the crystallisation. The homologue **NF4** revealed the shortest temperature range of the $N_F$ phase and crystallised at about 74°C. On the other hand, the longest homologues **NF5** and **NF6** did not crystallise during the DSC measurements at all. For these homologues, the $N_F$ phase persisted during the second and third cooling-heating DSC cycles. The stability of the $N_F$ phase for these two homologues was confirmed during electro-optical measurements: the $N_F$ phase was stable for several hours at RT. The DSC thermograph for the homologue **NF6** is demonstrated in Fig. 2. For the first heating of the sample, the melting point (m.p.) was established; for the second heating run, the $N_F$ phase melted at a temperature corresponding to $T_{iso}$. A glassy transition was clearly distinguishable and its temperature, $T_g$, was determined from the onset calculated at a half heat capacity, $c_p$, see Table 1. Glassy properties and ability to form fibres from the melted compound **NF5** is demonstrated in Supplemental file (Fig. S2).

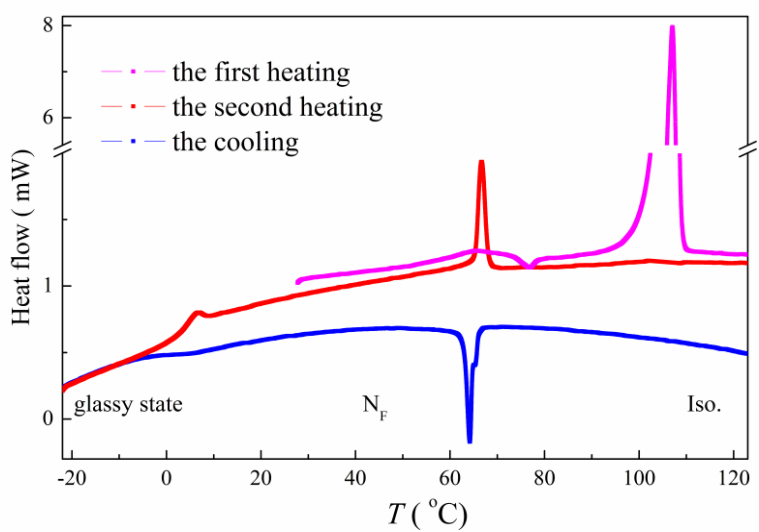

Fig. 2. DSC thermograph detected for **NF6** during the first and second heating and cooling runs.

Table 1. Calorimetric data taken from DSC measurements: melting point, m.p., detected at the first heating run, the $N_F$-Iso phase transition temperature, $T_{iso}$, and the glassy transition temperature, $T_g$. All temperatures are presented in °C, and enthalpy changes, ΔH, in J/g, are in square brackets at the corresponding temperatures.

|     | m.p. ΔH (J/g) | $T_{iso}$ (°C) ΔH (J/g) | $T_g$ ΔH (J/g) |
| --- | --- | --- | --- |
| NF1 | 188 [+98.6] | 170 [-2.73] | 24 [+0.47] |
| NF2 | 150 [+71.3] | 136 [-7.59] | 30 [+0.42] |
| NF3 | 156 [+73.8] | 116 [-7.13] | 15 [+0.27] |
| NF4 | 144 [+80.2] | 96 [-6.11] | - |
| NF5 | 120 [+55.1] | 82 [-4.86] | -9 [+0.44] |
| NF6 | 104 [+51.4] | 65 [-3.33] | 4 [+0.28] |

In the polarising microscope, we observed various textural features in different commercial or home-made cells. There are two basic geometries for rod-shaped liquid crystalline molecules: in HG cells, the molecules are oriented along the cell surface, and in the HT cell, a homeotropic anchoring ensured molecular orientation perpendicular to this direction. In the HG cell, the alignment is provided by rubbed polyimide layers with a small pretilt to arrange defect-free textures. The pretilt results in nonzero polar surface energy as was pointed out by Chen et al. [14]. Two kinds of HG cells were available, with parallel (HG-P) or antiparallel (HG-A) rubbing directions on opposite glass surfaces.

Let us start with HG-A cells and compare the results for various cell thicknesses. In this geometry, we observed two kinds of domains. The texture in 5μm HG-A cell for the studied homologue **NF6** is shown in Fig. 3. The dominating type of domains are twisted domains, which were described for the $N_F$ phase in literature [12]. The twisted domains are recognisable when slightly uncrossing the analyser from the crossed position. Another type of domains can be observed in less extensive areas of the HG-A samples. In the upper right part of Fig. 3, we found "red-colour" domains with characteristic borderline approximately parallel to the rubbing direction. The red colour was typical for these domains in 5 μm HG-A cell, see Fig. S2-S5 in Supplemental for other homologues. Extinction position in these domains are not easy to be established and the colour of these domains changes when rotating the sample with respect to the polariser position. We did not find twisted domains in HG-P cells with parallel alignment. In this geometry, we observed homogeneously aligned area as well as "red" domains, as is demonstrated for **NF6** in Fig. S6 in Supplemental file.

We concentrate on twisted domains, which are very frequent in the HG-A geometry. We found that for very thin HG-A sample, the twisted domain can be extended for a large area by quick cooling from the isotropic phase (with a rate >20 K/min). Rather big twisted domains separated by a zig-zag borderline are demonstrated in Fig. 4(a) for **NF6** in 1.6 μm HG-A cell. One can see that the borderline between the twisted domains is oriented approximately perpendicularly to the rubbing direction. When we turn the analyser from the crossed position by an angle of ~ 20 degrees, we clearly observe two kinds of domains (see insets in Fig. 4); the sense of twist is opposite for two neighbouring domains and they are separated by 2π disclination line. In the paper by Sebastian [12], similar domains were observed for another type of ferroelectric nematogen and designated "sierra-domains". In our particular case, these twisted domains reveal sharper contour and can be renamed as "shark-domains". For the homologue **NF5**, the twisted domains are demonstrated in Fig. S4 in Supplemental. Schematic picture of molecular twist between surfaces with antiparallel alignment is shown in Fig. 4(b).

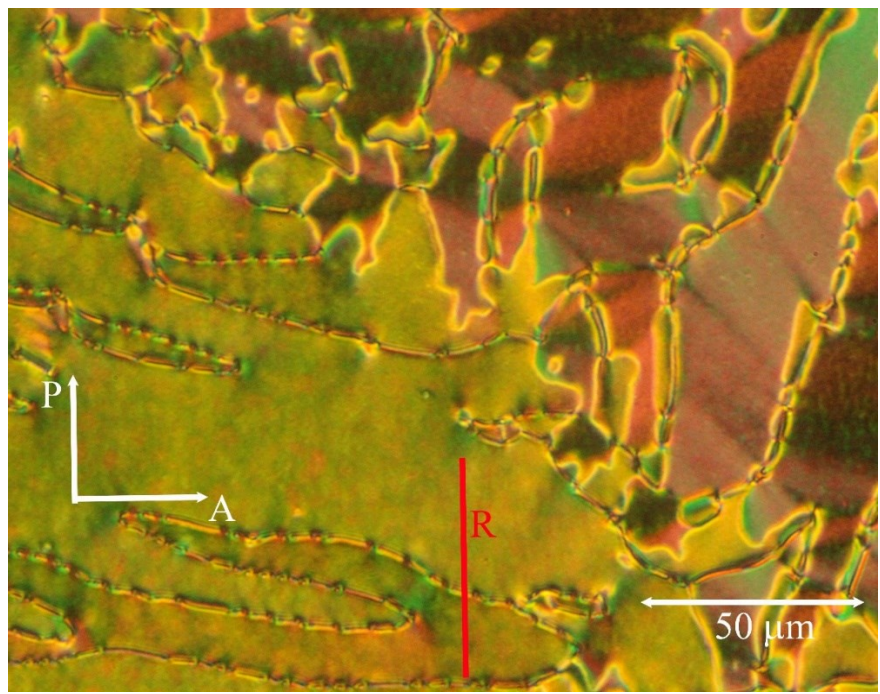

Fig. 3. Microphotograph of **NF6** homologue in 5 μm HG-A cell. The red arrow (R) marks the rubbing direction, the white arrows show the polariser (P) / analyser (A) directions.

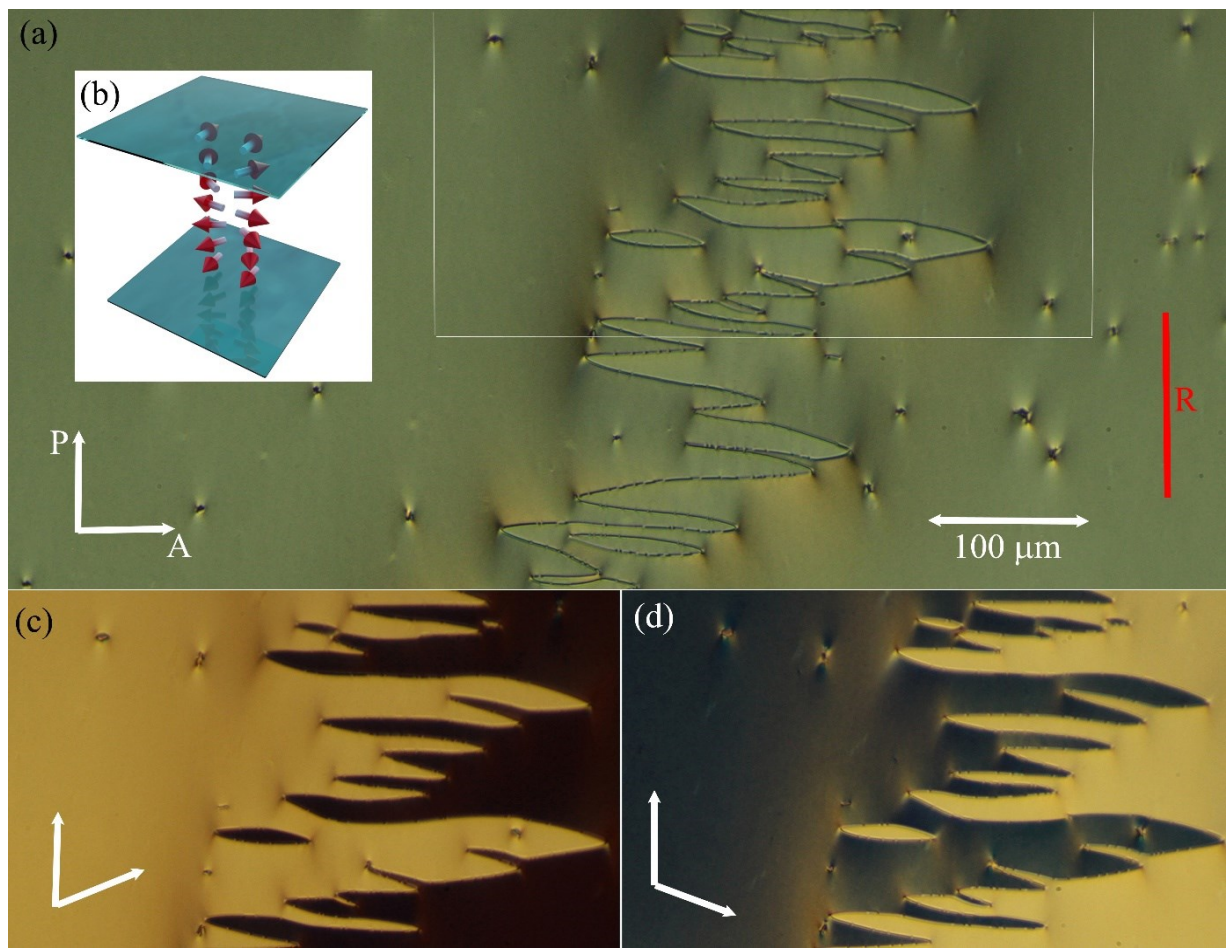

Fig. 4.　　Textures of **NF6** in 1.6 μm HG-A cell under a polarizing microscope (a) between crossed polarisers, the red arrow marks the rubbing direction, R, the orientation of the analyser (A) and the polariser (P) is schematically shown by white arrows. In the figure (b) there is a schematic arrangement of molecules in neighbouring twisted domains between glass surfaces with antiparallel rubbing. The part of the figure (a) marked by white lines is shown in (c) and (d) when A is rotated by an angle of about 20 degrees counterclockwise or clockwise from the crossed position.

An application of an electric field in HG geometry led to a rather complex effect. The colour of twisted domains slightly changed under the applied electric field and additional stripes appeared across the twisted domain structure approximately parallel to the rubbing direction. As the application of the field in the HG cell supplied only limited information and a detail analysis is rather problematic, we studied HT cells under applied bias. In this geometry, the applied electric field is approximately parallel to the molecular dipole moment and we can observe a rearrangement of molecules. In Fig. 5, we demonstrate the HT texture for homologue **NF6** with and without applied electric field of about 5 V/μm. In the upper part of Fig. 5, an area without electrode is observed. When the field is switched on (Fig. 5(b)), the molecules reorient along the field and the texture under the electrode area becomes black. After switching the electric field off, the HT texture turns back to a lighter type, similar to the virgin texture (Fig. 5(a)), within several seconds.

We investigated the switching properties of the studied compounds in HT geometry. Due to electrostatic interactions, the results are influenced both by the cell geometry and by the character of the aligning layer. For n=1-4, the homologues **NFn** reveal strong vitrification and an increase in viscosity when approaching the glass transition temperature $T_g$. This temperature is relatively high and the samples feature a higher conductivity, which limited our studies for these homologues. On the contrary, homologues **NF5** and **NF6** could be subjected to the applied field for a longer time (several hours); the polarisation was measured repeatedly and the results were reproducible. At the room temperature, these two homologues stay in LC phase for a long time and they start to crystallise only after several hours.

For homologue **NF5**, the temperature dependence of the polarisation is presented in Fig. 6(a). The polarisation values are calculated by the time-integration of a switching current profile. In Fig. 6(b), the switching current is plotted versus the applied electric field at a frequency 10 Hz and at temperature 52 °C. For both homologues **NF5** and **NF6**, we detected a continuous increase in polarisation values on cooling process in the $N_F$ phase. A coexistence of the Iso and $N_F$ phases was checked under the polarising microscope and it was observed only in a narrow temperature interval of about 2 °C. The decrease in polarisation values shown in Fig. 6(a) is connected with an increase in switching time. Such a slowing-down of molecular dynamics is connected with an increase in the sample viscosity.

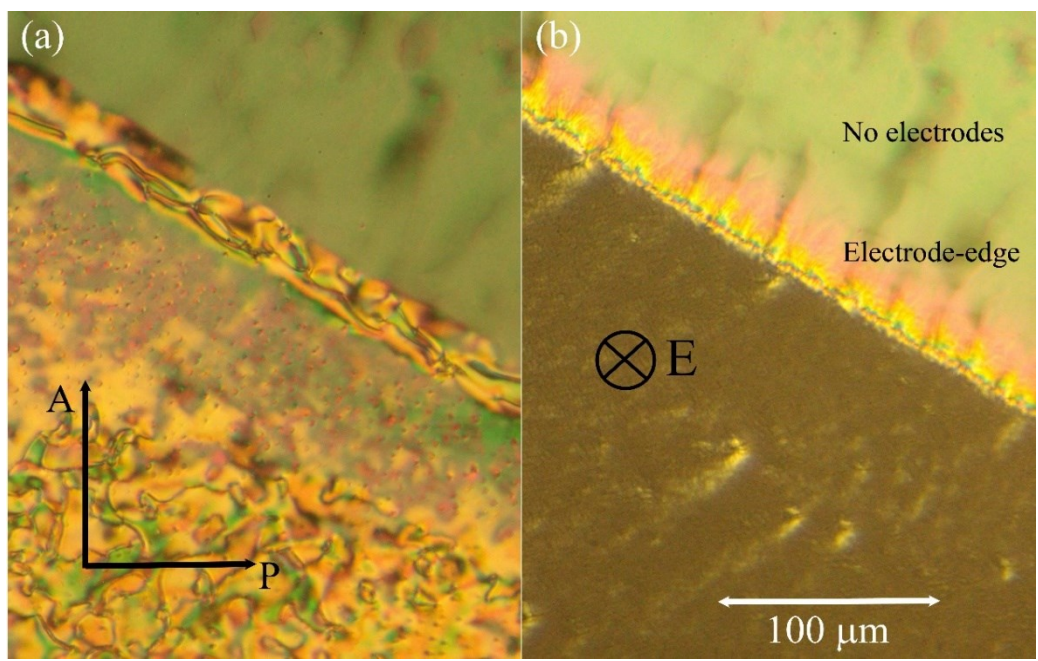

Fig. 5.    Texture of **NF6** in 5 μm thick HT cell, (a) without electric field and (b) under applied electric field of about 5 V/μm perpendicular to the cell. The orientation of polarisers and of the applied electric field are marked by black symbols for illustration. Upper part of the figure shows an area without electrodes.

To analyse the effect of the applied field in different cell geometries, we prepared a home-made gap-cell with in-plane electrodes. Two glass slides were separated by copper 35 μm thick ribbons, with a gap distance of about 1 mm. In this cell, the domains disappeared under the applied electric field as all the molecules were aligned along the applied electric field. After the switching-off of the external electric field, the domain structure was partially reconstructed in several seconds. Microphotographs can be found in Supplemental file (Fig. S8). Unfortunately, the thickness of 35 μm was rather large to reach homogeneous alignment through the whole cell thickness. Additionally, we are aware that the applied electric field was not homogeneously distributed. Technological tasks of the cell preparation and detailed analysis of the defects in electric field are still under work.

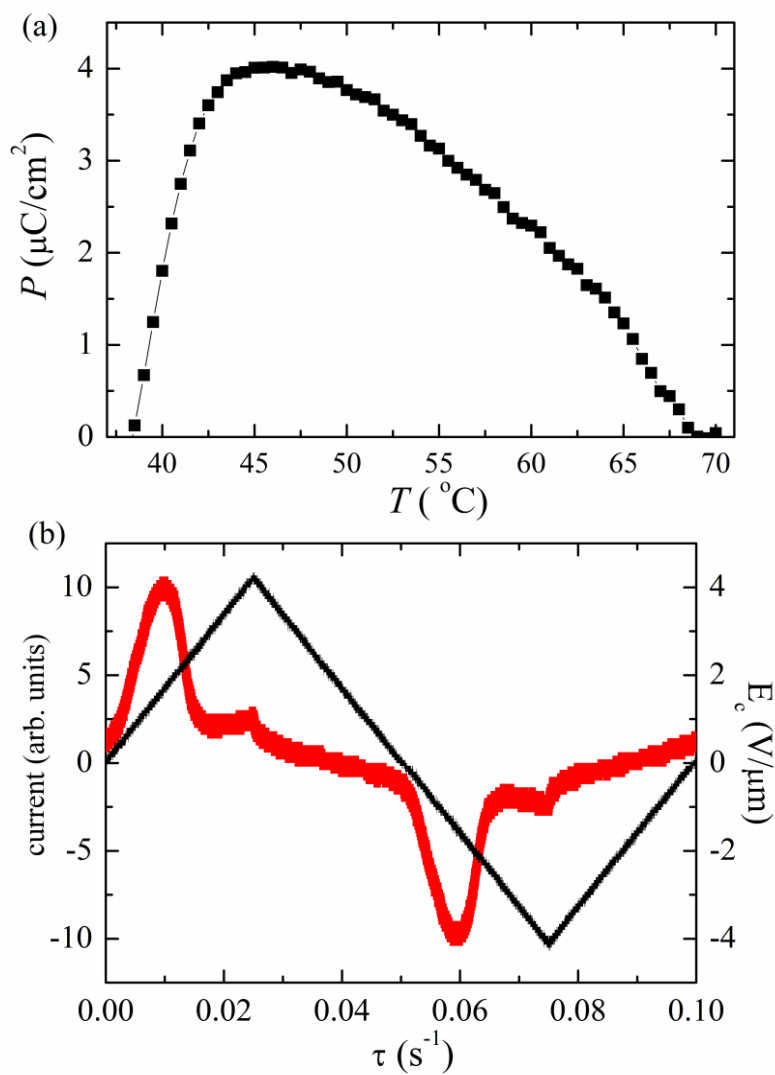

Fig. 6. (a) The temperature dependence of the polarisation of **NF5**, which was calculated from the polarisation current. (b) The current profile at a temperature T=52 °C is demonstrated with a triangular profile of the applied electric field at a frequency 10 Hz.

We measured the dielectric spectra of all compounds in a temperature range from the isotropic liquid to RT in order to study the molecular dynamics. The applied measuring field was smaller than 0.01 V/μm (higher probing fields could influence the dielectric measurements as the studied compounds are really sensitive to external fields). In the ferroelectric $N_F$ phase, we found one distinct quite strong relaxation mode appearing at the Iso-$N_F$ phase transition on cooling and remaining visible down to RT. On the other hand, when the sample is in the crystalline state, this mode is not present and the permittivity is low (<10). We demonstrate three-dimensional plots of the real, $\varepsilon'$, and imaginary, $\varepsilon''$, parts of permittivity versus frequency and temperature, $T$, for compound **NF5** in Fig. 7. For homologues **NF2**, **NF3** and **NF6**, the 3D-plots of permittivity are shown in Supplemental file, Figs. S9-S11. All the presented dielectric data were obtained in 12 μm thick cells with gold electrodes and no surfactant layers.

We encountered a disturbing effect of surfactant, similarly as it was mentioned in previous works dealing with dielectric spectroscopy of the $N_F$ phase [16]. For such a type of polar phase, it was reported that the polymer layers effectively influence the permittivity measurements. Due to a non-conductive character of polymer layers on the cell surfaces, there is a barrier which causes a spatial variation of the charge and influences the measured effective permittivity values. We fitted the dielectric data to the Cole-Cole formula (see Supplemental file for the details) to obtain information about the dielectric strength, $\Delta\varepsilon$, and the relaxation frequency, $f_r$. We detected large only slightly temperature dependent values of $\Delta\varepsilon$ up to $15\times10^3$. In contrast, the relaxation frequency decreases within the whole temperature range of the $N_F$ phase on cooling and follows the Arrhenius law. Such behaviour is documented in Fig. 8 for homologue **NF6**, which followed Arrhenius behaviour ideally and the activation energy, $E_a$, was calculated to be 102 kJ/mol. For other compounds, the linearity of $f_r$ in logarithmic scale (versus 1/T in absolute temperature scale) was confirmed only far from the Iso-$N_F$ phase transition (see Fig. S12 in Supplemental file). Non-homogeneity of molecular alignment and/or influence of electrodes should be taken into consideration to explain the deviation from Arrhenius law.

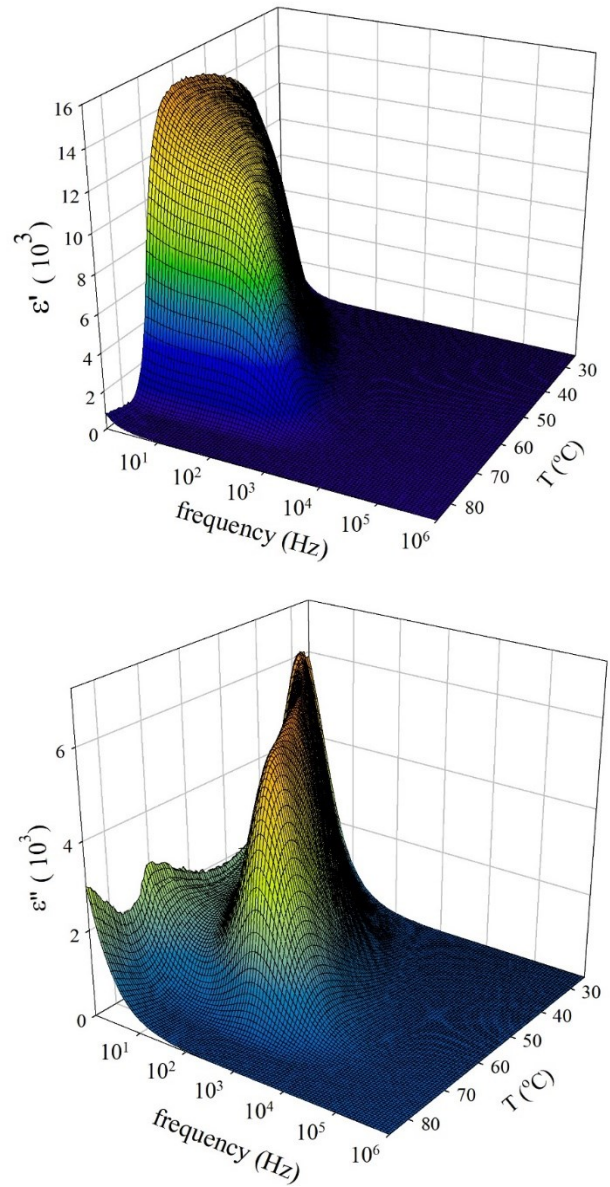

Fig. 7. 3D-plot of (a) real, ε', and (b) imaginary, ε'', parts of the permittivity versus the frequency and the temperature, $T$, for compound **NF5**. Dielectric measurements were performed in 12 μm cell with gold electrodes and no surfactant layer.

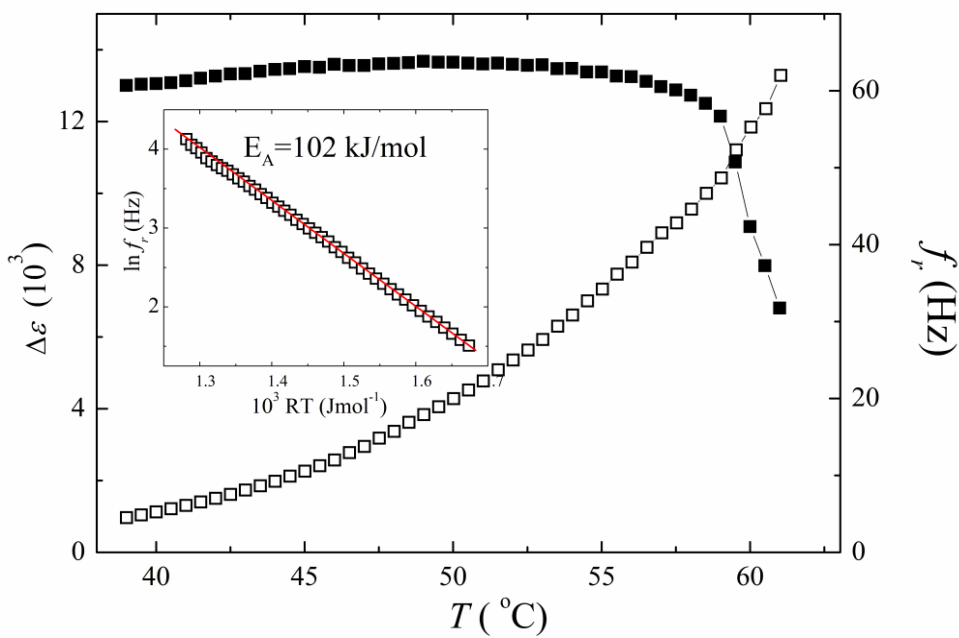

Fig. 8.    Temperature dependences of (a) the dielectric strength, Δε, and relaxation frequency, $f_r$, for **NF6** in 12 μm cell without surfactant layer. In the inset $f_r$ is presented in the logarithmic scale versus reciprocal temperature, 1/T, in Kelvins and the activation energy, $E_A$, was established from the slope.

Strong polar character of the $N_F$ phase was proved by SHG measurements. The SHG experiments were carried out in transmission configuration according the scheme described in Supplemental file. We utilised HG cells and SHG measurement results are presented for compounds **NF5** and **NF6** in Fig. 9. On cooling the sample from the isotropic phase, the SHG signal abruptly grows from zero value at the transition temperature to $N_F$ phase. With ongoing temperature decrease, the SHG intensity slows down its increase, reaches the maximum and slowly starts to decrease. All of this happens within the $N_F$ phase, where we would expect a gradual increase in the SHG signal upon cooling. Moreover, even for the weakest applied intensity of the fundamental laser beam, a small drop in SHG intensity was detected in subsequent measuring runs at the same temperature. From this it follows that the decrease in the SHG signal upon cooling may be explained by partial decomposition of our samples caused by rather strong intensity of the pulse laser beam.

X-ray scattering experiments confirmed nematic character of the observed mesophase. Nematic phase is characterised by the long-range orientational order and only broad diffuse peaks of low intensity can be detected. For homologue **NF5**, the signal at small scattering angles is rather wide and can be fitted with two signals, with maxima corresponding to 18.8 Å and 10.5 Å at T=75°C, 22.5 Å and 10.4 Å at T=30°C. As the length of molecules, $l$, can be approximately established as $l$~20.9 Å, the peak at the small scattering angle matches perfectly to the long dimension of the molecules. The peak at a wide-angle region has also a very broad profile with the maximum corresponding to 4.4 Å for all measuring temperatures, and it corresponds to an average distance between the molecules.

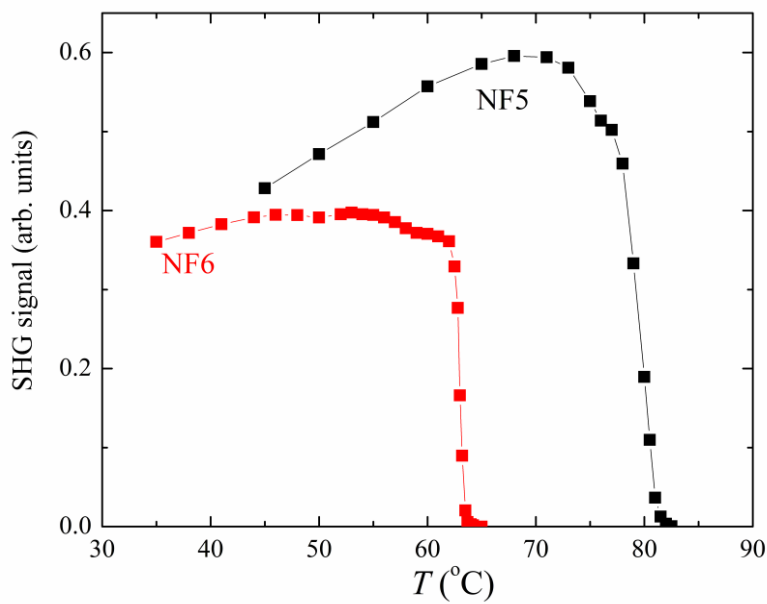

Fig. 9. SHG signals for **NF5** and **NF6** in HG cells

## 4. Conclusions

We proposed a new structural modification of highly polar molecules self-assembling and forming the ferroelectric nematic phase $N_F$. All the prepared compounds exhibit a direct phase transition to the ferroelectric nematic phase on cooling from the isotropic phase. In the presented homologue series, a prolongation of a side-chain resulted in the $N_F$ phase persistence down to the room temperatures and stability for at least several hours. Ferroelectric character of the nematic phase was proven by several experimental techniques. Characteristic textural features for the ferroelectric nematics were observed in several sample geometries. The ferroelectric switching process was detected and the polarisation was calculated from the measured polarisation current. The values of polarisation were found to increase continuously on cooling from the isotropic phase, reaching up to 4 µC/cm$^2$. For all the studied homologues, the dielectric studies show a strong polar mode characteristic for the $N_F$ phase, disappearing in the isotropic or crystalline phases. The dielectric strength of this mode exceeds values of about $15\times10^3$, which is the maximum reached for the $N_F$ phase up to now. Nevertheless, the characterisation of the defects in the $N_F$ phase and the role of the electrodes are not yet completely solved and will need a deeper insight. The dipole moment of the molecules was calculated and established to be about 14 D, which is larger than the value reported for DIO or RM734.

The discovery of ferroelectricity for nematics opened new opportunities in the liquid crystal research and generally in the field of condensed matter. The $N_F$ phase represents a highly polar structure responsive to very small applied fields and it features a variety of new effects induced by the confining surfaces. Generally, the application potential of the $N_F$ phase is immense and not yet completely explored. Our particular room-temperature-stable soft phase exhibits huge dielectric constant and can be important in future for the development of memory devices, capacitors and actuators.

**Disclosure statement**

No potential conflict of interest was reported by the authors.

**Acknowledgments**

Authors acknowledge project MAGNELIQ, that received funding from the European Union's Horizon 2020 research and innovation programme under grant agreement No 899285; and project 22-16499S from the Czech Science Foundation. V.N. is grateful to Damian Pociecha and Ewa Gorecka from Warsaw University for their help with x-ray measurements.

# Supplemental information

# Dimethylamino terminated ferroelectric nematogens revealing high permittivity

Martin Cigl, Natalia Podoliak, Tomáš Landovský, Dalibor Repček, Petr Kužel, and Vladimíra Novotná*

*Institute of Physics of the Czech Academy of Sciences, Na Slovance 2, Prague, Czech Republic*

**Contents**

1. **Syntheses and compound characterisation**

1.1. **General synthesis**

All starting materials and reagents were purchased from Sigma-Aldrich, Acros Organics or Lach:Ner. All solvents used for the synthesis were "p.a." grade. Tetrahydrofuran was further distilled from calcium hydride to obtain sufficiently dry solvent. $^1$H NMR spectra were recorded on Varian VNMRS300 instrument; deuteriochloroform ($CDCl_3$) and hexadeuteriodimethyl sulfoxide (DMSO-$d_6$) were used as solvents and the signals of the solvent served as an internal standard. Chemical shifts (δ) are given in ppm and *J* values are given in Hz. Elemental analyses were carried out on Elementar vario EL III instrument. The purity of all final compounds was checked by HPLC analysis (high-pressure pump ECOM Alpha; column WATREX Biospher Si 100, 250 × 4 mm, 5 μm; detector WATREX UVD 250) and were found to be >99.8 %. Column chromatography was carried out using Merck Kieselgel 60 (60–100 μm).

Synthesis of materials started from commercial 4-aminosalicylic acid (**1**, see Scheme 1). Its amino group was protected by acetylation and the carboxylic group was protected by alkylative esterification by methyl iodide, so as neither of the two groups interfere with the alkylation of phenolic hydroxyl. Protected derivative **2** was then alkylated by 1-bromoalkanes to get a series of alkyl homologues **3-n**. In the next steps, the acetyl group was cleaved by acidic hydrolysis under mild conditions and the liberated amino group was alkylated by dimethyl sulphate yielding the key intermediate, acid **4-n**. The lowest alkyl homologue (**4-1**) was synthesised directly from acid **1** by alkylation with the excess of dimethyl sulphate. The second part of the molecular core was synthesised from 4-hydroxybenzoic acid (**5**), which was protected by the reaction with 3,4-dihydro-2H-pyrane and reacted with 4-nitrophenol in a DCC-mediated esterification. The protected hydroxyl group was then liberated by the treatment with *p*-toluenesulfonic acid. The final step of the synthesis was esterification of acids **4-n** with phenol **6** mediated by EDC.

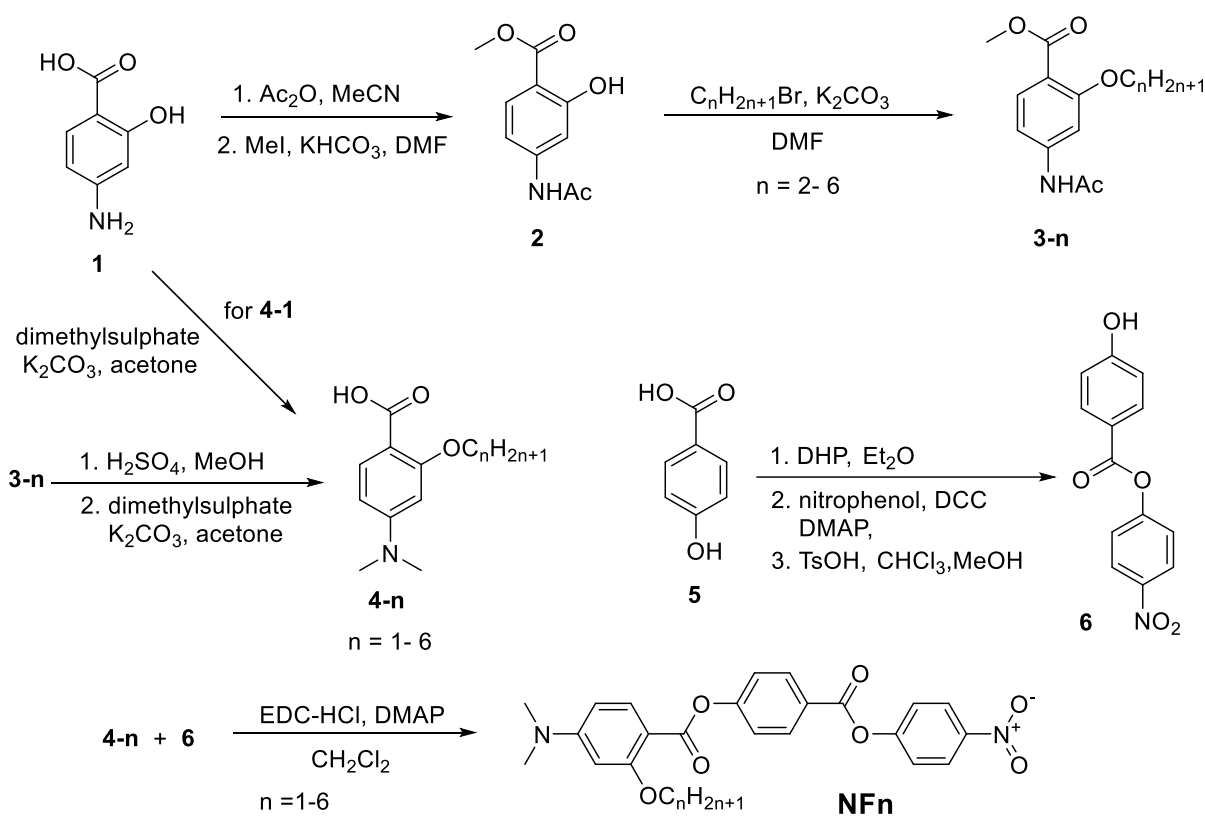

Scheme 1.    Synthetic procedures for the preparation of target compounds **NFn**.

## 1.2.    Synthetic procedures

*Methyl 4-acetamido-2-hydroxybenzoate* (**2**)
Acetic hydride (30 mL, 0.31 mol) was added dropwise to the suspension of powdered 4-aminosalicylic acid (20.0 g, 0.13 mol) in acetonitrile (250 mL). The reaction mixture was stirred for 2 h and the resulting suspension filtered. The filter cake was washed by the small amount of acetonitrile to remove the residue of acetic acid and dried in a vacuum dryer at 40 °C. Yield 24.77 g (96 %).

Dry 4-acetamidosalicylic acid (24.0 g, 0.12 mol) was dissolved in DMF. Powdered $KHCO_3$ was added with stirring, resulting in $CO_2$ evolution. Then methyl iodide was added dropwise and the reaction mixture stirred for 6 h under anhydrous conditions ($CaCl_2$ tube). The resulting suspension was poured into water and neutralised with concentrated HCl. White precipitate was filtered off and crystallised from 50% aqueous methanol. Yield 24.43 g (95 %). $^1$H NMR (DMSO-$d_6$) δ: 10.23 (1 H, s), 7.71 (1 H, d, *J*=8.8 Hz), 7.37 (1 H, d, *J*=1.8 Hz), 7.05 (1 H, dd, *J*=8.8, 2.3 Hz), 3.85 (3 H, s), 2.07 (3 H, s).

*General procedure for alkylation of benzoate* **2**

Benzoate 2 was dissolved in dry DMF and powdered $K_2CO_3$ and KI (omitted if iodoalkane was used) were added with stirring. Mixture was heated to 50 °C and 1-bromoalkane was added. Reaction was stirred at 50 °C under anhydrous conditions ($CaCl_2$ tube) for 10 h. The cooled resulting mixture was poured into cold water, neutralised with concentrated HCl and the precipitated product was filtered off and crystallised from ethanol.

*Methyl 4-acetamido-2-ethoxybenzoate* (**3-1**)
The reaction of benzoate **2** (5.0 g, 23.90 mmol) with ethyl iodide (5.45 g, 39.32 mmol) in the presence of $K_2CO_3$ (5.0 g, 36.18 mmol) in dry DMF (50 mL) yielded 4.83 g (85 %) of **3a**. $^1$H NMR (CDCl$_3$) δ: 11.25 (1 H, br. s.), 8.36 (1 H, d, *J*=2.3 Hz), 7.94 (1 H, d, *J*=9.4 Hz), 6.58 (1 H, dd, *J*=9.1, 2.6 Hz), 4.11 (2 H, q, *J*=7.0 Hz), 3.89 (3 H, s), 2.23 (3 H, s), 1.42 (3 H, t, *J*=7.0 Hz).

*Methyl 4-acetamido-2-propoxybenzoate* (**3-2**)
The reaction of benzoate **2** (5.0 g, 23.90 mmol) with 1-bromopropane (8.71 g, 70.82 mmol) in the presence of $K_2CO_3$ (10.0 g, 72.36 mmol) in dry DMF (60 mL) yielded 3.53 g (58 %) of **3b**. $^1$H NMR (CDCl$_3$) δ:10.95 (1 H, s), 7.79 (1 H, d, *J*=8.8 Hz), 7.60 (1 H, d, *J*=2.3 Hz), 6.81 (1 H, dd, *J*=8.8, 2.3 Hz), 4.09 (2 H, t, *J*=6.5 Hz), 1.81 - 2.05 (2 H, m), 1.10 (3 H, t, *J*=7.3 Hz).

*Methyl 4-acetamido-2-butoxybenzoate* (**3-c**)
The reaction of benzoate **2** (5.0 g, 23.90 mmol) with 1-bromobutane (4.53 g, 32.40 mmol) in the presence of $K_2CO_3$ (5.0 g, 36.18 mmol) in dry DMF (50 mL) yielded 5.42 g (86 %) of **3c**. $^1$H NMR (CHLOROFORM-*d*) δ ppm 10.85 (1 H, d, *J*=8.8 Hz), 7.77 (1 H, s), 7.60 (1 H, d, *J*=2.3 Hz 6.79 (1 H, dd, *J*=8.8, 2.3 Hz), 4.04 (2 H, t, *J*=6.5 Hz), 3.86 (2 H, s), 2.19 (3 H, s), 1.72 - 1.93 (2 H, m), 1.42 - 1.61 (2 H, m), 0.97 (3 H, t, *J*=7.3 Hz).

*Methyl 4-acetamido-2-(pentyloxy)benzoate* (**3-d**)
The reaction of benzoate **2** (10.0 g, 47.80 mmol) with 1-iodopentane (19.88 g, 98.37 mmol) in the presence of $K_2CO_3$ (15.0 g, 0.11 mol) in dry DMF (100 mL) yielded 9.81 g (75 %) of **3d**. $^1$H NMR (CDCl$_3$) δ:10.86 (1 H, s), 7.80 (1 H, d, *J*=8.8 Hz), 7.58 (1 H, d, *J*=2.3 Hz), 6.79 (1 H, dd, *J*=8.8, 2.3 Hz), 4.01 (2 H, t, *J*=6.7 Hz), 1.77 - 2.03 (2 H, m), 1.31 - 1.57 (4 H, m), 0.91 (3 H, t, *J*=7.3 Hz).

*Methyl 4-acetamido-2-(hexyloxy)benzoate* (**3-e**)
The reaction of benzoate **2** (10.0 g, 47.80 mmol) with 1-bromohexane (15.78 g, 95.60 mmol) in the presence of $K_2CO_3$ (15.0 g, 0.11 mol) in dry DMF (100 mL) yielded 9.73 g (69 %) of **3e**. $^1$H NMR (CDCl$_3$) δ: 10.83 (1 H, s), 7.79 (1 H, d, *J*=8.8 Hz), 7.59 (1 H, d, *J*=2.3 Hz), 6.80 (1 H, dd, *J*=8.8, 2.3 Hz), 4.02 (2 H, t, *J*=6.7 Hz), 3.86 (3 H, s), 2.19 (3 H, s), 1.71 - 1.91 (2 H, m), 1.42 - 1.56 (2 H, m), 1.20 - 1.41 (4 H, m), 0.91 (3 H, t, *J*=7.3 Hz).

*General procedure for deacetylation of amino group*

Methyl 4-acetamido-2-(alkoxy)benzoate **3** was dissolved in methanol at 50 °C and the concentrated H$_2$SO$_4$ was carefully added dropwise. The reaction mixture was stirred at 50 °C for 30 min and then poured into cold water and neutralised with NaOH. The neutral dispersion of the product in water was extracted with ethyl acetate, combined organic layers were washed with water and brine. After drying with anhydrous MgSO$_4$, the solvent was removed on rotary evaporator and the residue was purified by column chromatography on silica gel to yield methyl 2-(alkoxy)-4-aminobenzoate as an intermediate.
*NOTE*: The presence of water in the deacetylation reaction (e.g. use of diluted H$_2$SO$_4$) leads to considerable amounts of decarboxylation byproduct.

*General procedure for methylation of amino group*

Methyl 2-(alkoxy)-4-aminobenzoate was dissolved in DMSO and powdered K$_2$CO$_3$ was added with stirring. A mixture was heated to 50 °C and dimethyl sulphate was added dropwise. The reaction mixture was stirred at the same temperature and under anhydrous conditions (CaCl$_2$ tube) overnight. The progress of the reaction was monitored using TLC (CH$_2$Cl$_2$-acetone 95 : 5). The resulting mixture was filtered and solid Na$_2$S was added to the filtrate. The mixture was stirred for 4 h at room temperature and then poured into water. After 30 min of standing, the solution was neutralised with the concentrated acetic acid and the precipitated product was collected by filtration. A crude product was purified by column chromatography on silica gel and crystallised from methanol.

*4-(Dimethylamino)-2-methoxybenzoic acid* (**4-1**)
Compound **4a** was synthesised by direct methylation of acid **1** using the general procedure for methylation of amino group described above. Reaction of benzoic acid **1** (10.0 g, 65.30 mmol) with dimethyl sulphate (42.45 g, 0.33 mol) in the presence of K$_2$CO$_3$ (50.0 g, 0.36 mol) in DMSO (150 mL) and subsequent treatment with Na$_2$S (16.0 g, 0.21 mol) yielded 10.01 g (79 %) of **4-1**.

*4-(Dimethylamino)-2-ethoxybenzoic acid* (**4-2**)
Following the general procedure, starting from methyl 4-acetamido-2-ethoxybenzoate **3-1** (4.80 g, 20.23 mmol), which was deacylated using H$_2$SO$_4$ (5.0 mL, 96%) in methanol (50 mL). The free amine was methylated using dimethyl sulphate (10.52 g, 80.90 mmol) and K$_2$CO$_3$ (12.0 g, 86.82 mmol) in DMSO (50 mL) followed by the treatment with Na$_2$S (1.60 g, 20.50 mmol) yielded 2.29 g (54 %) of **4-2**. $^1$H NMR (CDCl$_3$) δ: 10.71 (1 H, br. s.), 7.99 (1 H, d, *J*=9.4 Hz), 6.37 (1 H, dd, *J*=8.8, 2.3 Hz), 6.10 (1 H, d, *J*=2.3 Hz), 4.29 (2 H, q, *J*=7.0 Hz), 3.05 (6 H, s), 1.55 (3 H, t, *J*=7.0 Hz).

*4-(Dimethylamino)-2-propoxybenzoic acid* (**4-3**)
Using the described general procedure: 4-acetamido-2-propoxybenzoate **3-2** (3.50 g, 13.93 mmol) was deacylated using H$_2$SO$_4$ (3.5 mL, 96%) in methanol (40 mL). Liberated amine was methylated using dimethyl sulphate (7.24 g, 55.68 mmol) and K$_2$CO$_3$ (7.70 g, 55.71 mmol) in DMSO (50 mL) followed by the treatment with Na$_2$S (1.10 g, 14.09 mmol) yielded 1.91 g (61 %) of **4-3**. $^1$H NMR (CDCl$_3$) δ: 10.73 (1 H, br. s.), 8.01 (1 H, d, *J*=9.4 Hz), 6.42 (1 H, dd, *J*=9.1, 2.1 Hz), 6.19 (1 H, d, *J*=1.8 Hz), 4.19 (2 H, t, *J*=6.5 Hz), 3.07 (6 H, s), 1.95 (2 H, sext., 7.2 Hz), 1.10 (3 H, t, *J*=7.3 Hz)

*2-Butoxy-4-(dimethylamino)benzoic acid* (**4-4**)
Using the general deacetylation and alkylation protocol: 4-acetamido-2-butoxybenzoate **3-3** (5.30 g, 19.98 mmol) was deacylated using H$_2$SO$_4$ (5.0 mL, 96%) in methanol (50 mL). Liberated amine was methylated using dimethyl sulphate (10.40 g, 79.98 mmol) and K$_2$CO$_3$ (11.50 g, 83.21 mmol) in DMSO (50 mL) followed by the treatment with Na$_2$S (1.60 g, 20.50 mmol) yielded 2.42 g (51 %) of **4-4**. $^1$H NMR (CDCl$_3$) δ: 10.71 (1 H, br. s.), 8.00 (1 H, d, *J*=8.8 Hz), 6.40 (1 H, dd, *J*=8.8, 2.3 Hz), 6.17 (1 H, d, *J*=2.3 Hz), 4.21 (2 H, t, *J*=6.7 Hz), 3.06 (6 H, s), 1.85 - 1.94 (2 H, m), 1.43 - 1.58 (2 H, m), 0.92 (3 H, , t, *J*=7.3 Hz).

*4-(Dimethylamino)-2-(pentyloxy)benzoic acid* (**4-5**)
Using the above mentioned general protocols: 4-acetamido-2-(pentyloxy)benzoate **3-4** (9.50 g, 34.76 mmol) was deacylated using H$_2$SO$_4$ (5.0 mL, 96%) in methanol (150 mL). Liberated amine was methylated using dimethyl sulphate (27.12 g, 0.21 mol) and K$_2$CO$_3$ (29.0 g, 0.21 mol) in DMSO (150 mL) followed by the treatment with Na$_2$S (10.90 g, 0.14 mol) yielded 5.41 g (62 %) of **4-3**. $^1$H NMR (CDCl$_3$) δ: 10.70 (1 H, br. s.), 8.00 (1 H, d, *J*=8.8 Hz), 6.41 (1 H, dd, *J*=8.8, 2.3 Hz), 6.18 (1 H, d, *J*=2.3 Hz), 4.21 (2 H, t, *J*=6.7 Hz), 3.06 (6 H, s), 1.82 - 1.99 (2 H, m), 1.31 - 1.55 (4 H, m), 0.94 (3 H, , t, *J*=7.3 Hz).

*4-(Dimethylamino)-2-(hexyloxy)benzoic acid* (**4-6**)
Using the above mentioned general protocols: 4-acetamido-2-(hexyloxy)benzoate **3-3** (9.50 g, 32.38 mmol) was deacylated using H$_2$SO$_4$ (5.0 mL, 96%) in methanol (150 mL). Liberated amine was methylated using dimethyl sulphate (21.10 g, 0.16 mol) and K$_2$CO$_3$ (23.0 g, 0.17 mol) in DMSO (150 mL) followed by the treatment with Na$_2$S (7.20 g, 0.10 mol) yielded 4.82 g (56 %) of **4-6**. $^1$H NMR (CDCl$_3$) δ: 10.72 (1 H, br. s.), 7.99 (1 H, d, *J*=8.8 Hz), 6.37 (1 H, dd, *J*=8.8, 2.3 Hz), 6.11 (1 H, d, *J*=2.3 Hz), 4.20 (2 H, t, *J*=6.7 Hz), 3.06 (6 H, s), 1.85 - 1.99 (2 H, m), 1.21 - 1.51 (6 H, m), 0.91 (3 H, , t, *J*=7.3 Hz).

*4-Nitrophenyl 4-hydroxybenzoate* (**6**)
3,4-Dihydro-2*H*-pyrane (14.30 g, 0.17 mol) was added dropwise to the suspension of 4-hydroxybenzoic acid (13.80 g, 0.10 mol) in diethylether (200 ml). The reaction mixture was stirred overnight under anhydrous conditions (CaCl$_2$ tube) and then filtered. The filter cake contained the majority of desired 4-((tetrahydro-2H-pyran-2-yl)oxy)benzoic acid. The filtrate was vigorously stirred with aqueous NaOH (80 ml, 10%) for 30 min, and then the aqueous layer was separated and neutralised by HCl. The pH was further adjusted to ca. 4 using acetic acid. The precipitated solid was collected, washed with cold water and dried under vacuum and finally combined with the dry portion obtained from the filter cake.

4-((Tetrahydro-2*H*-pyran-2-yl)oxy)benzoic acid (31.35 g, 0.14 mol) and 4-nitrophenol (19.60 g, 0.14 mol) were dissolved in dry THF (250 mL) and cooled to ca. 10 °C. Then *N,N′*-dicyclohexylcarbodiimide (DCC, 30.60 g, 0.15 mol ) and 4-(dimethylamino)-pyridine (DMAP, 5.60 g, 46.22 mmol) were added and the reaction mixture was stirred under anhydrous conditions for 12 h. The precipitated *N,N′*-dicyclohexylurea was filtered off and the filtrate diluted with ethyl acetate (100 mL). The resulting solution was washed with diluted HCl (100 mL, 1 : 15), then with water and the solvents were removed on rotary evaporator. The solid residue was dissolved in CHCl$_3$-methanol mixture (1 : 1) and toluenesulfonic acid (4.0 g, 23.22 mmol) was added. The reaction mixture was stirred at 45 °C for 1 h and then evaporated to dryness on rotary evaporator. A crude product was crystallised from acetone. Yield 28.71 g (66 %). $^1$H NMR (DMSO-d$_6$) δ: 8.33 (2 H, d, *J*=8.6 Hz), 8.08 (2 H, d, *J*=8.8 Hz), 7.58 (2 H, d, *J*=8.6 Hz), 7.20 (2 H, d, *J*=8.8 Hz), 5.62 – 5.65 (1 H, m), 3.63 - 3.75 (1 H, m), 3.48 - 3.60 (1 H, m), 1.21 - 1.97 (6 H, m).

*General procedure for EDC-mediated esterification*

2-(Alkoxy)-4-(dimethylamino)benzoic acid **4-n** and 4-nitrophenyl 4-hydroxybenzoate (**6**) were suspended in dry dichloromethane (50 ml) and cooled to 2 – 8 °C in ice-water bath. Then *N*-(3-dimethylaminopropyl)-*N′*-ethylcarbodiimide hydrochloride (EDC) and 4-(*N,N*-dimethylamino)pyridine (DMAP) (0.1 g, 0.82 mmol) were added. The reaction mixture was stirred for 2 hours under anhydrous conditions and the temperature was let rise as ice in the cooling bath melted. The resulting solution diluted with CH$_2$Cl$_2$ and washed with water and brine. Organic layer was dried over anhydrous magnesium sulphate and evaporated on the rotary evaporator. The residue was purified by column chromatography on silica gel in CH$_2$Cl$_2$-acetone eluent and recrystallised from acetone.

*4-[(4-Nitrophenoxy)carbonyl]phenyl 4-(dimethylamino)-2-methoxybenzoate* (**NF1**)
4-(Dimethylamino)-2-methoxybenzoic acid (**4-1**, 78.1 mg, 0.40 mmol) was esterified with 4-nitrophenyl 4-hydroxybenzoate (**6**, 104.5 mg, 0.40 mmol) using EDC (81 mg, 0.42 mmol) and DMAP (51.0 mg, 0.42 mmol) in dichloromethane (2.0 mL) as described in general procedure. Yield 92.5 mg (53 %). $^1$H NMR (CDCl$_3$) δ: 8.33 (2 H, d, *J*=8.8 Hz), 8.24 (2 H, d, *J*=8.8 Hz), 8.03 (1 H, d, *J*=9.4 Hz), 7.41 (4 H, dd, *J*=14.1, 8.8 Hz), 6.33 (1 H, dd, *J*=8.8, 2.3 Hz), 6.16 (1 H, d, *J*=2.3 Hz), 3.95 (3 H, s), 3.11 (6 H, s). Anal. calcd. for C$_{23}$H$_{20}$N$_2$O$_7$: C 63.30, H 4.62, N 6.42; found C 63.86, H 4.68, N 6.47 %.

*4-[(4-Nitrophenoxy)carbonyl]phenyl 4-(dimethylamino)-2-ethoxybenzoate* (**NF2**)
4-(Dimethylamino)-2-methoxybenzoic acid (**4-1**)
The reaction of 4-(dimethylamino)-2-ethoxybenzoic acid (**4-2**, 1.0 g, 4.78 mmol) was esterified with 4-nitrophenyl 4-hydroxybenzoate (**6**, 1.24 g, 4.78 mmol) using EDC (1.0 g, 5.16 mmol) and DMAP (0.29 g, 2.39 mmol) in dichloromethane (30 mL) yielded 1.03 g (48 %). $^1$H NMR (CDCl$_3$) δ: 8.33 (2 H, d, *J*=9.4 Hz), 8.24 (2 H, d, *J*=8.2 Hz), 8.01 (1 H, d, *J*=9.4 Hz), 7.33 - 7.48 (4 H, m), 6.32 (1 H, dd, *J*=8.8, 2.3 Hz), 6.16 (1 H, d, *J*=1.8 Hz), 4.15 (2 H, d, *J*=7.0 Hz), 3.08 (6 H, s), 1.49 (3 H, t, *J*=7.0 Hz). $^{13}$C{H} NMR (CDCl$_3$) δ: 163.72 (s), 163.10 (s), 162.29 (s), 156.46 (s), 155.73 (s), 155.24 (s), 145.31 (s), 134.35 (s), 131.79 (s), 125.24 (s), 125.03 (s), 122.67 (s), 122.62 (s), 104.59 (s), 103.90 (s), 95.65 (s), 64.41 (s), 40.11 (s), 14.77 (s). Anal. calcd. for C$_{24}$H$_{22}$N$_2$O$_7$: C 64.00, H 4.92, N 6.11; found C 63.87, H 4.98, N 6.11 %.

*4-[(4-Nitrophenoxy)carbonyl]phenyl 4-(dimethylamino)-2-propoxybenzoate* (**NF3**)
Starting from 4-(dimethylamino)-2-propoxybenzoic acid (**4-3**, 1.25 g, 5.78 mmol) and 4-nitrophenyl 4-hydroxybenzoate (**6**, 1.50 g, 5.78 mmol) with EDC (1.18 g, 6.03 mmol) and DMAP (0.68 g, 5.61 mmol) in dichloromethane (50 mL) yielded 1.36 g (51 %). $^1$H NMR (CDCl$_3$) δ: 8.34 (2 H, d, *J*=8.8 Hz), 8.24 (2 H, d, *J*=8.8 Hz), 8.00 (1 H, d, *J*=9.4 Hz), 7.33 - 7.50 (4 H, m), 6.32 (1 H, dd, *J*=8.8, 2.3 Hz), 6.15 (1 H, d, *J*=2.3 Hz), 4.04 (2 H, t, *J*=6.5 Hz), 3.09 (6 H, s), 1.81 - 1.98 (2 H, m), 1.08 (3 H, t, *J*=7.3 Hz). $^{13}$C{H} NMR (CDCl$_3$) δ: 163.73 (s), 163.29 (s), 162.34 (s), 156.54 (s), 155.74 (s), 155.27 (s), 145.36 (s), 134.46 (s), 131.84 (s), 125.26 (s), 125.04 (s), 122.65 (s), 122.60 (s), 104.66 (s), 103.84 (s), 95.45 (s), 70.18 (s), 40.11 (s), 22.64 (s), 10.68 (s). Anal. calcd. for C$_{25}$H$_{24}$N$_2$O$_7$: C 64.65, H 5.21, N 6.03; found C 64.56, H 5.19, N 5.98 %.

*4-[(4-Nitrophenoxy)carbonyl]phenyl 2-butoxy-4-(dimethylamino)benzoate* (**NF4**)
Esterification of 2-butoxy-4-(dimethylamino)benzoic acid (**4-4**, 2.0 g, 8.43 mmol) with 4-nitrophenyl 4-hydroxybenzoate (**6**, 2.50 g, 9.64 mmol) using EDC (2.0 g, 10.22 mmol) and DMAP (0.58 g, 4.78 mmol) in dichloromethane (70 mL) yielded 2.31 g (51 %). $^1$H NMR (CDCl$_3$) δ: 8.34 (2 H, d, *J*=9.4 Hz), 8.24 (2 H, d, *J*=8.2 Hz), 8.00 (1 H, d, *J*=8.8 Hz), 7.31 - 7.50 (4 H, m), 6.32 (1 H, dd, *J*=8.8, 2.3 Hz), 6.16 (1 H, d, *J*=1.8 Hz), 4.08 (2 H, t, *J*=6.5 Hz), 3.09 (6 H, s), 1.74 - 1.96 (2 H, m), 1.42 - 1.66 (2 H, m), 0.95 (3 H, t, *J*=7.3 Hz). $^{13}$C{H} NMR (CDCl$_3$) δ: 163.73 (s), 163.33 (s), 162.30 (s), 156.52 (s), 155.74 (s), 155.25 (s), 145.34 (s), 134.47 (s), 131.82 (s), 125.25 (s), 125.02 (s), 122.66 (s), 122.60 (s), 104.65 (s), 103.83 (s), 95.43 (s), 68.35 (s), 40.15 (s), 31.30 (s), 19.25 (s), 13.85 (s). Anal. calcd. for C$_{26}$H$_{26}$N$_2$O$_7$: C 65.26, H 5.48, N 5.85; found C 65.15, H 5.16, N 5.80 %.

*4-[(4-Nitrophenoxy)carbonyl]phenyl 4-(dimethylamino)-2-(pentyloxy)benzoate* (**NF5**)
Following the general procedure above 4-(dimethylamino)-2-propoxybenzoic acid (**4-3**, 2.0 g, 7.96 mmol) and 4-nitrophenyl 4-hydroxybenzoate (**6**, 2.06 g, 7.94 mmol) were reacted in the presence of EDC (1.68 g, 8.59 mmol) and DMAP (0.50 g, 4.13 mmol) in dichloromethane (70 mL) yielded 1.76 g (45 %). $^1$H NMR (CDCl$_3$) δ: 8.33 (2 H, d, *J*=9.2 Hz), 8.24 (2 H, d, *J*=8.6 Hz), 7.99 (1 H, d, *J*=9.2 Hz), 7.32 - 7.49 (4 H, m), 6.32 (1 H, dd, *J*=8.9, 2.3 Hz), 6.15 (1 H, d, *J*=2.0 Hz), 4.07 (2 H, t, *J*=6.6 Hz), 3.08 (6 H, s), 1.79 - 1.94 (2 H, m), 1.26 - 1.55 (4 H, m), 0.83 - 0.94 (3 H, m). $^{13}$C{H} NMR (CDCl$_3$) δ: 163.73 (s), 163.33 (s), 162.30 (s), 156.52 (s), 155.74 (s), 155.25 (s), 145.34 (s), 134.47 (s), 131.82 (s), 125.25 (s), 125.02 (s), 122.66 (s), 122.60 (s), 104.65 (s), 103.83 (s), 95.43 (s), 68.67 (s), 40.13 (s), 28.92 (s), 28.17 (s), 22.42 (s), 14.00 (s). Anal. calcd. for C$_{27}$H$_{28}$N$_2$O$_7$: C 65.84, H 5.73, N 5.69; found C 65.59, H 5.78, N 5.65 %.

*4-[(4-Nitrophenoxy)carbonyl]phenyl 4-(dimethylamino)-2-(hexyloxy)benzoate* (**NF6**)
The reaction of 4-(dimethylamino)-2-ethoxybenzoic acid (**4-2**, 2.10 g, 7.91 mmol) was esterified with 4-nitrophenyl 4-hydroxybenzoate (**6**, 2.10 g, 8.10 mmol) using EDC (1.70 g, 8.69 mmol) and DMAP (0.96 g, 7.92 mmol) in dichloromethane (70 mL) yielded 1.63 g (41 %). $^1$H NMR (CDCl$_3$) δ: 8.32 (2 H, d, *J*=9.4 Hz), 8.23 (2 H, d, *J*=8.8 Hz), 7.99 (1 H, d, *J*=9.4 Hz), 7.31 - 7.52 (4 H, m), 6.32 (1 H, dd, *J*=8.8, 2.3 Hz), 6.15 (1 H, d, *J*=1.8 Hz), 4.07 (2 H, t, *J*=6.7 Hz), 1.77 - 1.93 (2 H, m), 1.41 - 1.57 (2 H, m), 1.18 - 1.39 (4 H, m), 0.76 - 0.94 (3 H, m). $^{13}$C{H} NMR (CDCl$_3$) δ: 163.70 (s), 163.32 (s), 162.27 (s), 156.54 (s), 155.74 (s), 155.25 (s), 145.35 (s), 134.46 (s), 131.77 (s), 125.23 (s), 125.01 (s), 122.64 (s), 122.56 (s), 104.69 (s), 103.85 (s), 95.48 (s), 68.70 (s), 40.07 (s), 31.52 (s), 29.20 (s), 25.69 (s), 22.53 (s), 13.99 (s). Anal. calcd. for C$_{28}$H$_{30}$N$_2$O$_7$: C 66.39, H 5.97, N 5.53; found C 66.21, H 5.90, N 5.49 %.

### 1.3. Equipment and apparatus

The compounds were studied by differential scanning calorimetry (DSC). Perkin-Elmer 7 Pyris calorimeter (Perkin Elmer, Shelton, CT, USA) was utilised and the measurements were conducted on cooling/heating runs at a rate of 10 K/min. The calorimeter was calibrated to the extrapolated onsets for the melting points of water, indium and zinc. A small amount of the studied compound (2-5 mg) was sealed into an aluminium pan and put into the calorimeter chamber. A nitrogen medium was utilised during the calorimetric measurements. The phase transition temperatures and the corresponding enthalpies were established from the second heating and the subsequent cooling runs.

Textures were observed under the polarising microscope Eclipse E600Pol (Nikon, Tokyo, Japan). We analysed the samples in various geometries. Two kinds of commercial cells were purchased with the thickness of 5 μm: HG cells with homogeneous anchoring (orienting molecules parallel to the cell surface) and HT cells with surfactant adjusting homeotropic arrangement of molecules (perpendicular to the surface). These cells consist of glasses with ITO transparent electrodes and materials were filled in the isotropic phase by capillary action. The Linkam E350 heating/cooling stage with TMS 93 temperature programmer (Linkam, Tadworth, UK) was utilised, with the temperature stabilisation within ±0.1 K.

The switching current profile versus time was detected by a digital oscilloscope Tektronix DPO4034 (Tektronix, Beaverton, OR, USA). Polarisation, P, was determined by the integration of the current profile when the electric field of triangular modulation at a frequency of 10 Hz was applied with the magnitude of 10 V/μm.

We measured the dielectric spectroscopy by Schlumberger 1260 impedance analyser (Schlumberger, Houston, TX, USA) and stabilised the temperature within ±0.1 K during the frequency sweeps in a range of 1 Hz ÷ 1 MHz. The permittivity, $\varepsilon*(f) = \varepsilon' - i\varepsilon''$, which is frequency dependent, was analysed with support of a modified version of the Cole-Cole formula:

$$\varepsilon^* - \varepsilon_\infty = \frac{\Delta\varepsilon}{1 + (if/f_r)^{(1-\alpha)}} - i\left(\frac{\sigma}{2\pi\varepsilon_0 f^n} + Af^m\right) \qquad (1),$$

where $f_r$ is the relaxation frequency, $\Delta\varepsilon$ is the dielectric strength, $\alpha$ is the distribution parameter of relaxation, $\varepsilon_0$ is the permittivity of vacuum, $\varepsilon_\infty$ is the high frequency permittivity, n, m, and A are the parameters of fitting. In formula (1) an ionic conductivity and ITO electrode effects were taken into consideration. The measured values of the real part of the permittivity, $\varepsilon'$, and the imaginary part, $\varepsilon''$, were simultaneously fitted to obtain the parameters $f_r$ and $\Delta\varepsilon$.

The polarisation current profile of electric field was detected by Tektronix DPO4034 digital oscilloscope (Tektronix, Oregon, US). The driving voltage from a generator (Agilent, California, US) was amplified by a linear amplifier providing the amplitude up to ±120 V.

The temperature-dependent second harmonic generation (SHG) measurements were conducted using an optical setup based on Ti:sapphire femtosecond laser (Spitfire ACE), which was amplified to produce 40 fs long pulses with 5 kHz repetition rate and central wavelength of 800 nm. For SHG we utilised HG cells and placed them into a Linkam stage, the temperature was stabilised with an accuracy ±0.1 K. The samples were illuminated by a collimated beam with pulses fluence of approximately 0.01 mJ/cm$^2$. The SHG signal generated in transmission configuration was appropriately filtered, then detected by an avalanche photodiode and amplified using a lock-in amplifier. The scheme of SHG measurements is shown in Figure S1.

For the x-ray studies, the Bruker D8 GADDS system was utilised: parallel CuK$\alpha$ beam formed by Goebel mirror monochromator, 0.5 mm collimator, modified Linkam heating stage, Vantec 2000 area detector. The samples for the diffraction experiments were prepared in a form of droplets on heated surface.

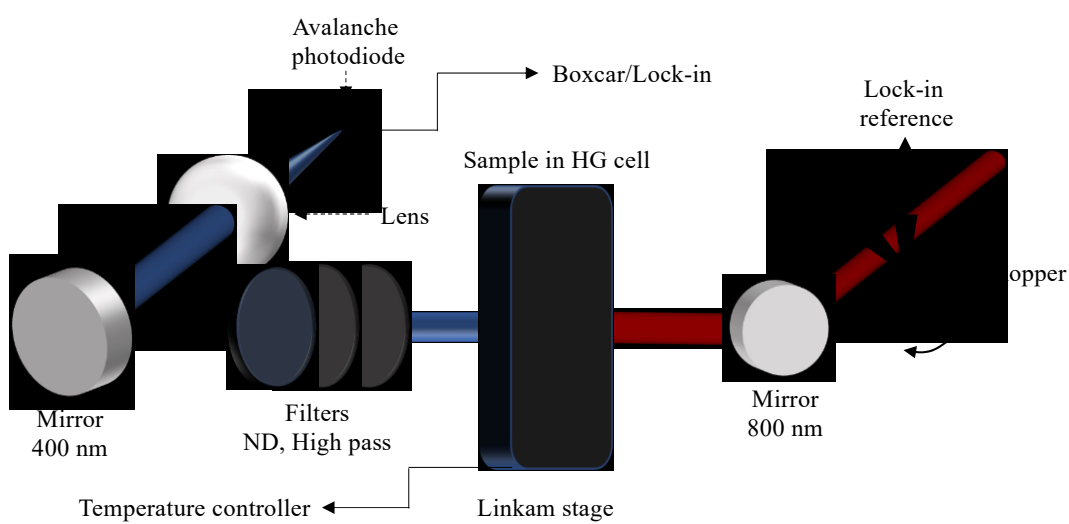

Figure S1.    SHG measurement scheme.

## 2.    Mesomorphic properties

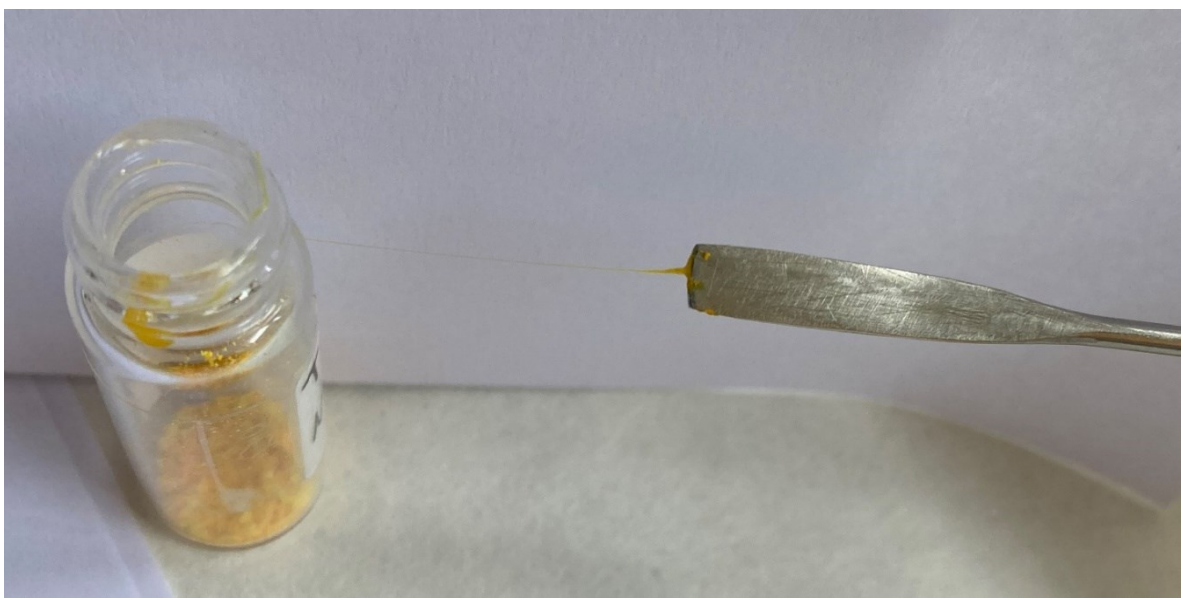

Figure S2. Vitrification process and creation of a fibre after melting of **NF5**.

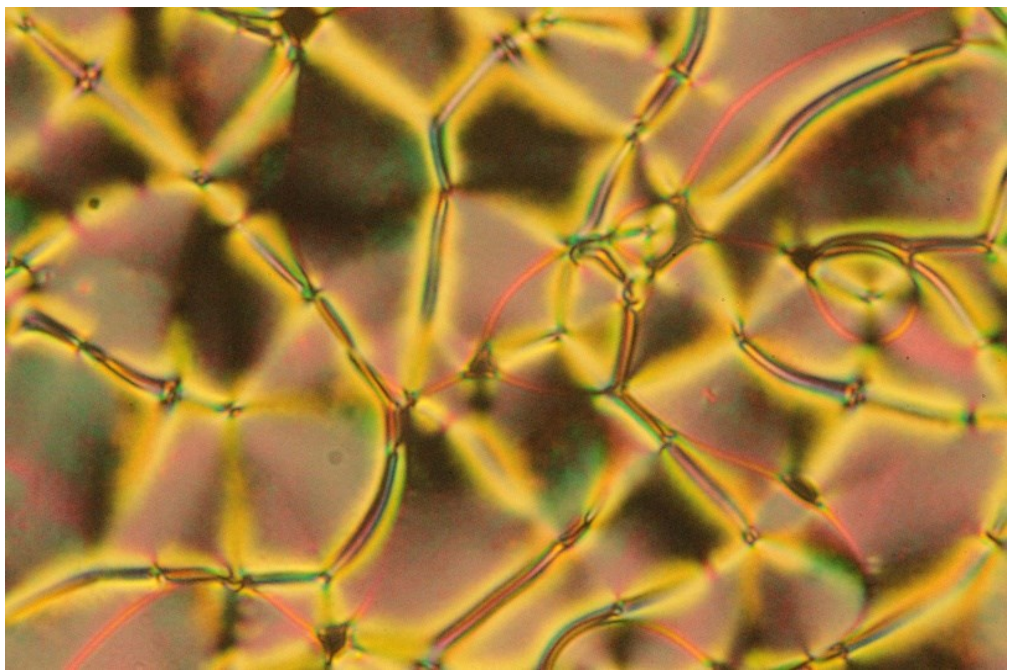

Figure S3. The microphotograph of the texture for homologue **NF2** detected in a 5 μm HG cell. The width of the photo corresponds to about 200μm.

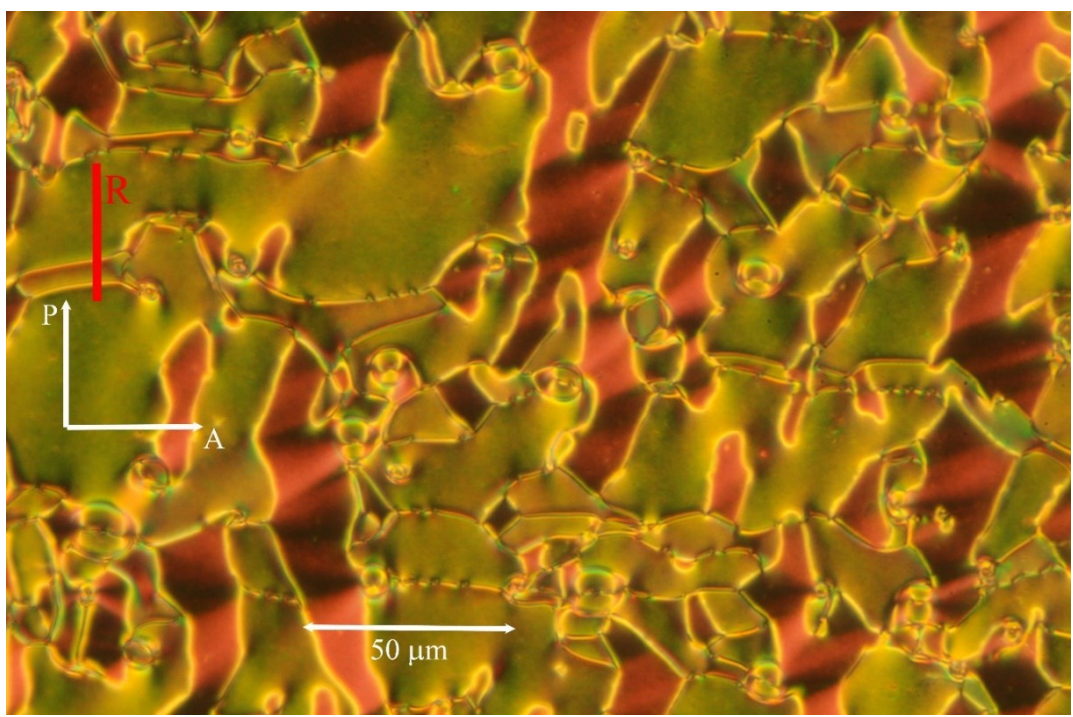

Figure S4. The microphotograph of **NF4** homologue in 5 μm HG-A cell. The Polariser orientation (white) and the rubbing direction (red) are marked.

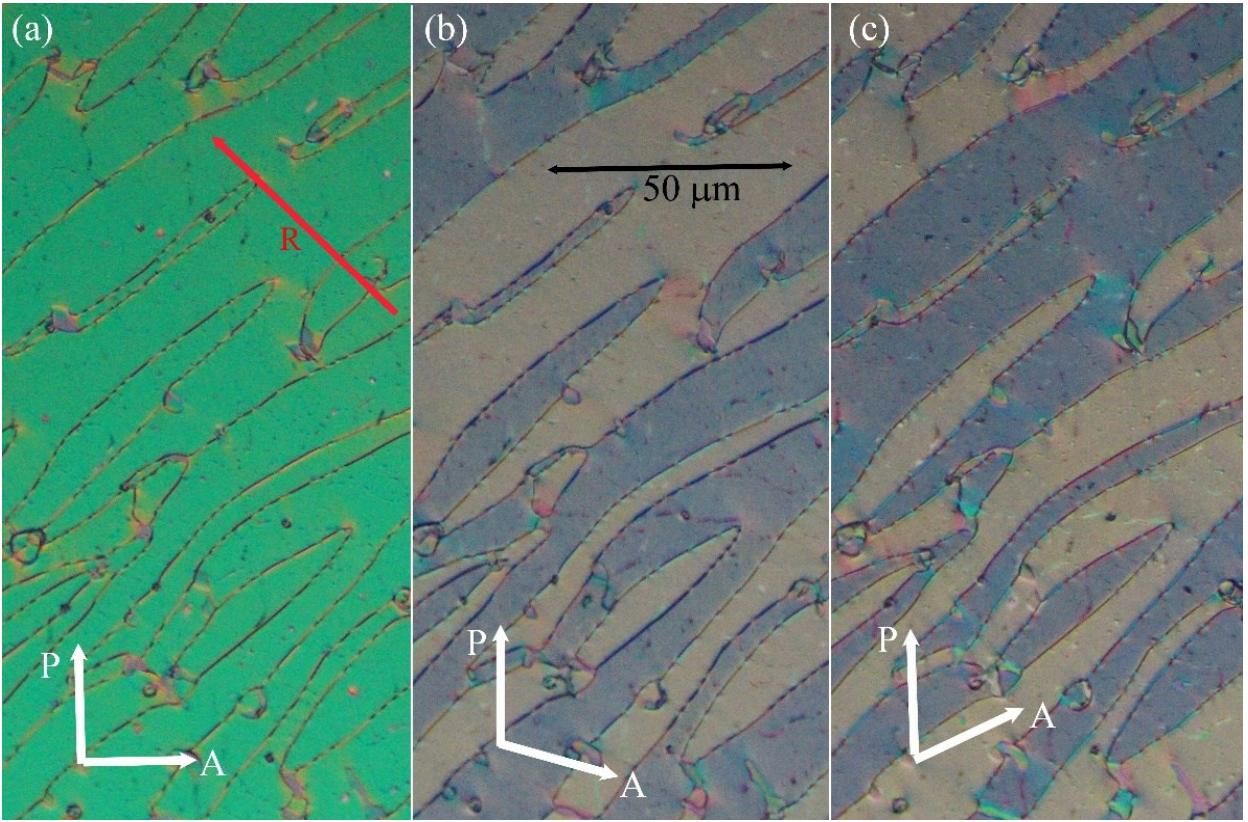

Figure S5   The texture of **NF5** in 5 μm HG-A cell under a microscope with (a) crossed polarisers, (b) and (c) with uncrossed position of polarisers (analyser rotated by an angle about 20 degrees). All figures show the same part of the sample; red arrow represents the rubbing direction and white arrows indicate the orientation of polarisers.

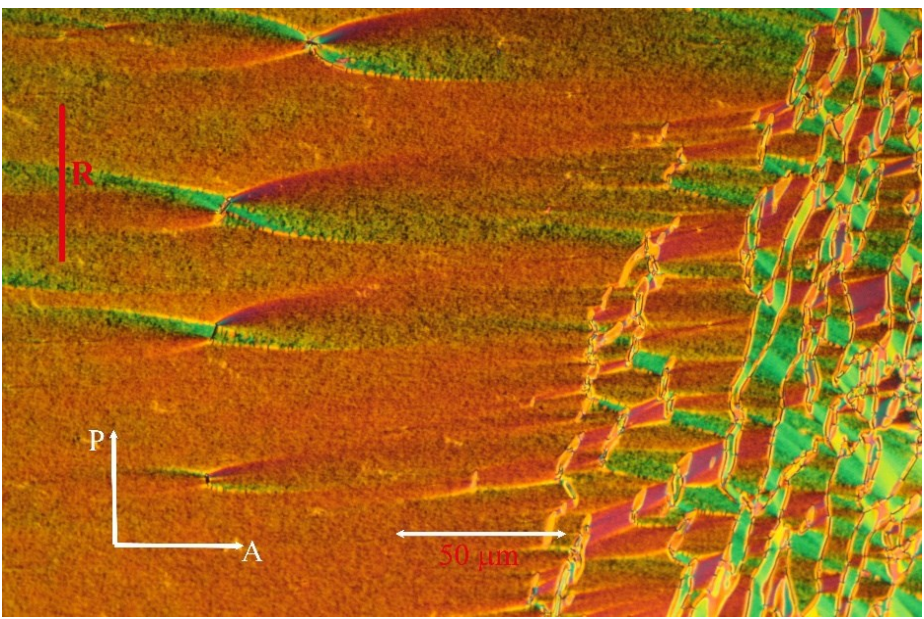

Figure S6.   The microphotograph of **NF6** homologue in 5 μm HG-P cell. The rubbing direction, R, is marked with a red line.

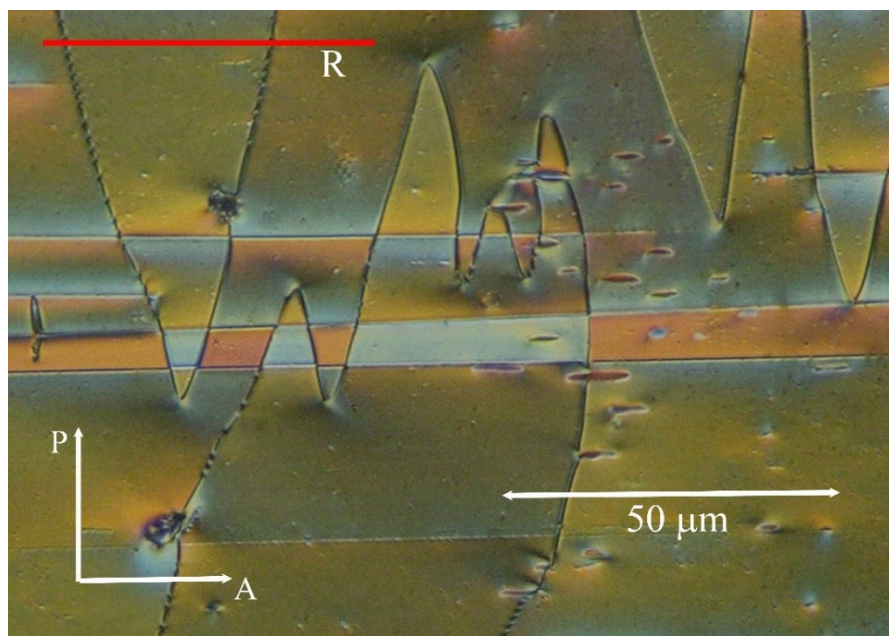

Figure S7. The photo of **NF6** homologue in 5 μm HG-A cell after the application of the external electric field of about 2 V/μm. The rubbing direction, R, is marked with a red line.

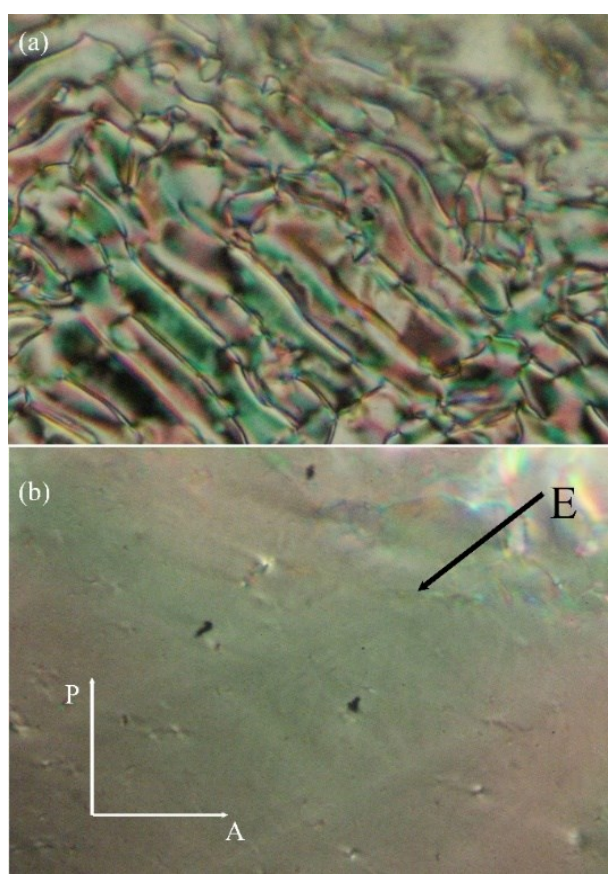

Figure S8. The texture of **NF6** in a special home-made gap-cell with a thickness of about 35 μm, defined by two copper electrodes. One electrode is located at the right upper corner out of the figure; the orientation of the applied electric field, E, is marked with the black arrow. For (a) no electric field was applied and for (b) the electric field of about 0.2 V/μm was applied.

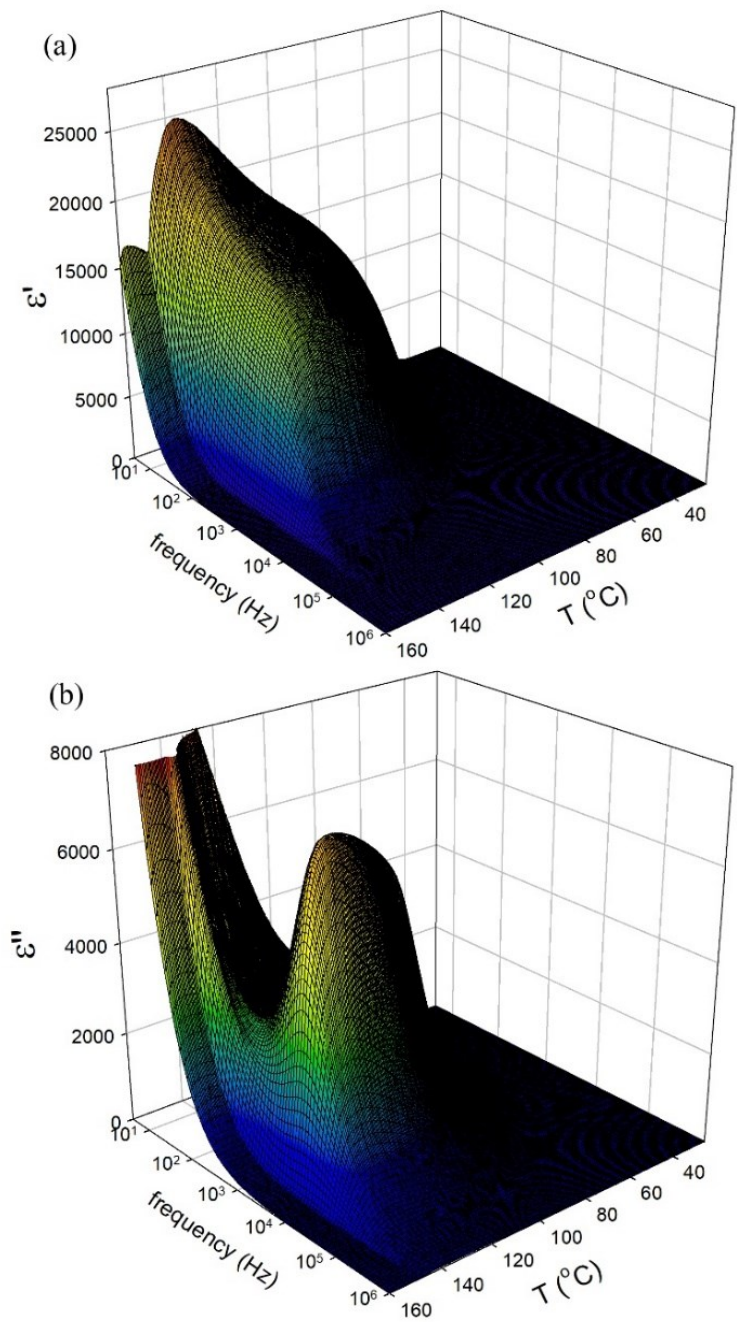

Figure S9. 3D-plot of (a) real, ε', and (b) imaginary, ε'', parts of the permittivity versus frequency and temperature, *T*, for compound **NF2**. Dielectric measurements were performed in 12 μm cell with gold electrodes and no surfactant layer.

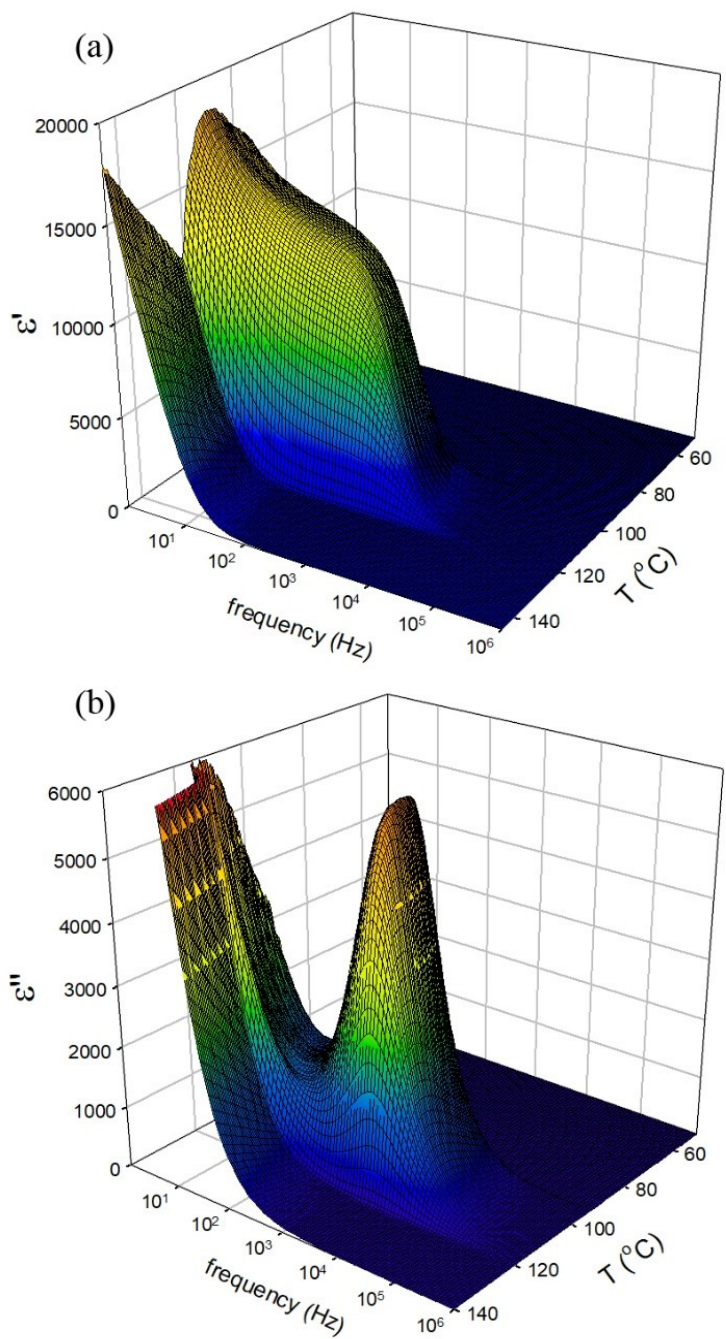

Figure S10.  3D-plot of (a) real, $\varepsilon'$, and (b) imaginary, $\varepsilon''$, parts of the permittivity versus frequency and temperature, $T$, for compound **NF3**. Dielectric measurements were performed in 12 μm cell with gold electrodes and no surfactant layer.

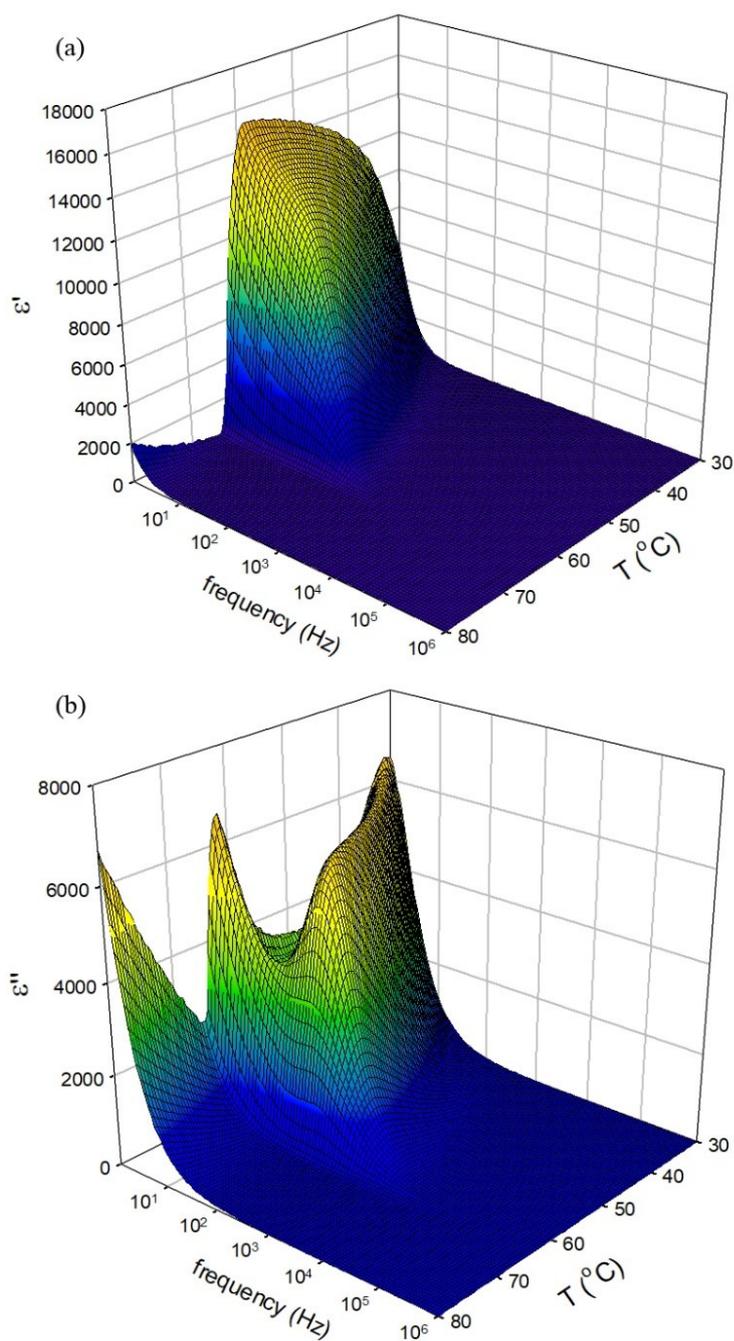

Figure S11.    3D-plot of (a) real, ε', and (b) imaginary, ε'', parts of the permittivity versus frequency and temperature, *T*, for compound **NF6**. Dielectric measurements were performed in 12 μm cell with gold electrodes and no surfactant layer.

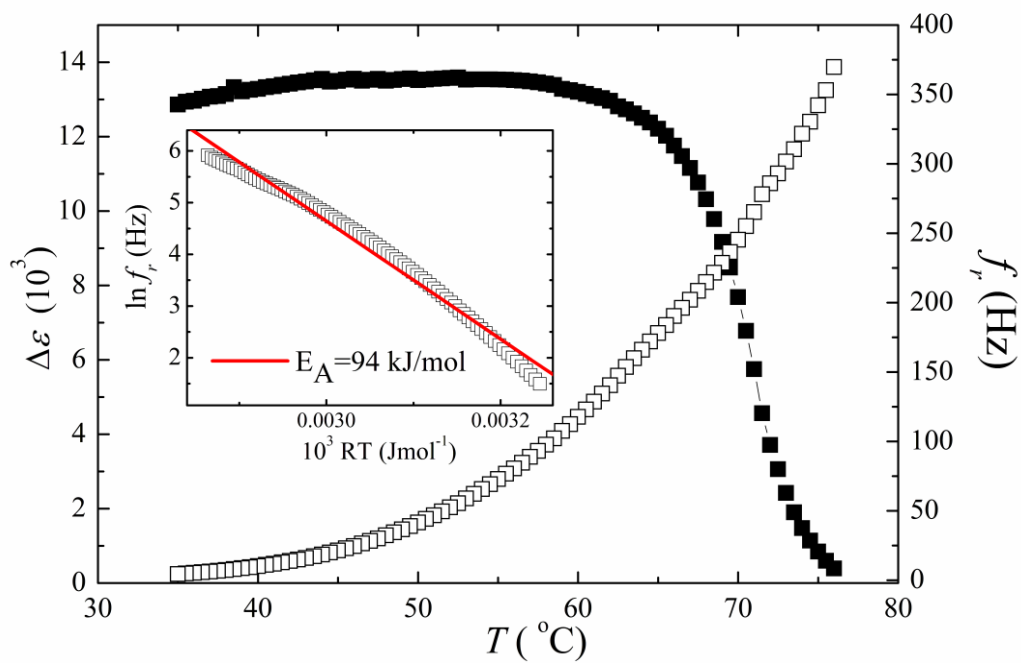

Figure S12. Temperature dependences of the dielectric strength, Δε, and the relaxation frequency, $f_r$, for **NF5** in 12 μm cell without surfactant layer. In the inset $f_r$ is presented in the logarithmic scale versus reciprocal temperature, 1/T, in Kelvins and the activation energy $E_A$ was established from the slope.

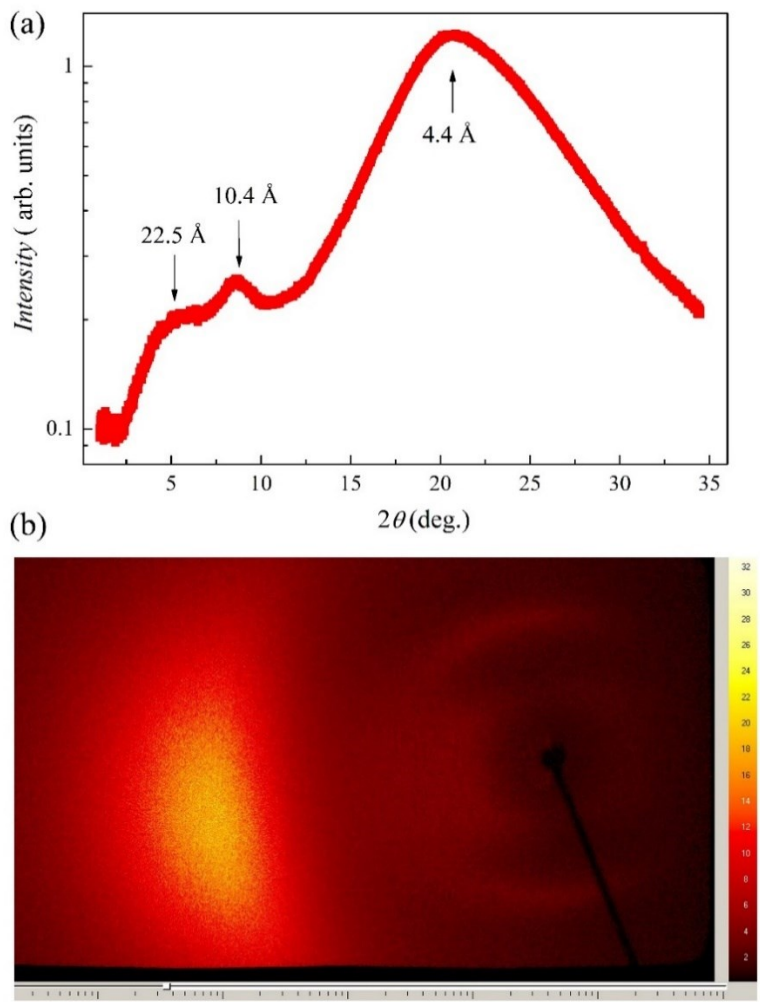

Figure S13. For **NF5** at the temperature T=30° C (a) the x-ray intensity versus the scattering angle, Θ,. (b) 2D pattern of the intensity at the same temperature. Scattering angles are in the logarithmic scale.

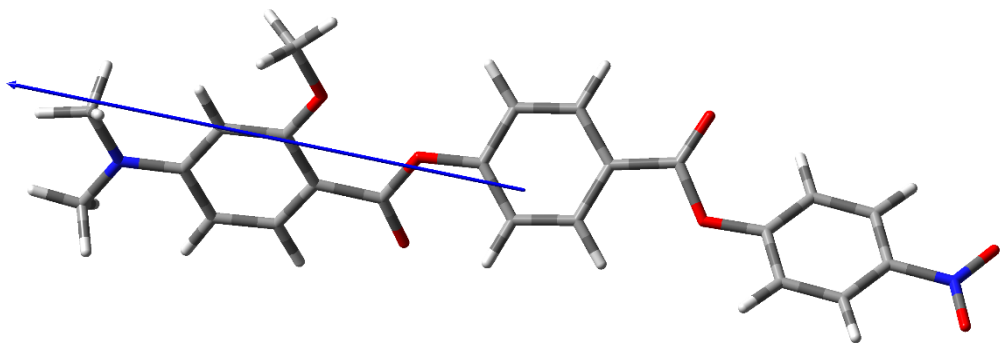

Figure S14. A model of **NF1** molecule with the orientation of the dipole moment (blue arrow).